\documentclass{article}

\usepackage{PRIMEarxiv}
\usepackage{booktabs}       % professional-quality tables
\usepackage{amsfonts}       % blackboard math symbols
\usepackage{nicefrac}
\usepackage{longtable}      % compact symbols for 1/2, etc.
\usepackage{microtype}      % microtypography
\usepackage{lipsum}
\usepackage{fancyhdr}       % header
\usepackage{graphicx} 
\usepackage{tabularx}
\usepackage{multirow}
\usepackage{multicol}
\usepackage{tabularray}
\usepackage{array}
\usepackage{caption}
\usepackage[
style=ieee,
]{biblatex}
\graphicspath{{media/}}     % organize your images and other figures under media/ folder

\title{Treatment and Follow-up Guidelines for Multiple Brain Metastases: A Systematic Review
}
\author{
  Ana Sofia Santos \\
  Department of Informatics, University of Minho, \\
  Braga, Portugal\\
  \And
  Matheus Silva  \\
  Director Radiotherapy Applications, Brainlab, \\
  Munich, Germany\\
  \And
  Crystian Saraiva  \\
  Medical Physicist/Radiotherapy, HCor,  \\
  São Paulo, Brazil\\
  \And
  José Miguel Soares  \\
  University of Minho, School of Medicine,\\
  MEng, PhD\\
  \And
  Victor Alves  \\
  Department of Informatics, University of Minho, \\
  Braga, Portugal\\
}

\begin{document}
\maketitle
\begin{abstract}
Brain metastases are a complication of primary cancer, representing the most common type of brain tumor in adults. The management of multiple brain metastases represents a clinical challenge worldwide in finding the optimal treatment for patients considering various individual aspects. Managing multiple metastases with stereotactic radiosurgery (SRS) is being increasingly used because of quality of life and neurocognitive preservation, which do not present such good outcomes when dealt with whole brain radiation therapy (WBRT). After treatment, analyzing the progression of the disease still represents a clinical issue, since it is difficult to determine a standard schedule for image acquisition. A solution could be the applying artificial intelligence, namely predictive models to forecast the incidence of new metastases in post-treatment images. Although there aren't many works on this subject, this could potentially bennefit medical professionals in early decision of the best treatment approaches.
\end{abstract}

% keywords can be removed

\section{Introduction}
\paragraph{}
Brain metastases, are a complication of primary cancer, representing the
most common type of brain tumor in adults \cite{Amsbaugh2022BrainMetastasis, 
Fox2011EpidemiologyTumors}. The improvement of imaging techniques has enabled 
earlier detection of smaller metastases therefore, the incidence of brain 
metastases has increased worldwide. A population-based study from Canada 
found that the annual average number of patients with BMs was over 3,200 at 
the time of diagnosis (data between the years of 2010 and 2017) 
\cite{Liu2022BrainIncidence}. 
An earlier population-based study conducted in Ontario estimated that the 
annual incidence of intracranial metastatic disease was over 3,500 (around 
24.2 among a population of 100,000 people) between the years of 2010 and 2018
\cite{Habbous2020IncidenceStudy}. In general, the incidence of brain 
metastases is not certainly
known, however, it is thought to be increasing \cite{BrainReview, 
Che2022RecentExperience, Liu2022BrainIncidence}. Brain metastases are 
secondary tumors consisting of cancer
cells that typically migrate through the bloodstream from the part of
the body where they originally started, to the brain. The majority of
patients who develop metastases to the brain have primary cancers
originating in the lung, breast, skin (melanoma), kidney, esophagus, and
colon/rectum \cite{Liu2022BrainIncidence, Habbous2020IncidenceStudy, 
Che2022RecentExperience, Singh2020EpidemiologyMetastases, 
Saha2013DemographicStudy}.
\paragraph{}
Because they occur in a very prompt proportion of patients, the
appropriate management of this disease is essential for improving
outcomes and quality of life in patients with advanced cancer.
Treatments for brain metastases range from several methods including
whole-brain radiation therapy (WBRT), stereotactic radiosurgery (SRS),
surgery resection, targeted therapies, and immunotherapies 
\cite{Suh2020CurrentMetastases, Lin2015TreatmentMetastases}. The optimal 
treatment for dealing with multiple metastases is an issue that lacks 
consensus, although several data is constantly released. 
\cite{Sittenfeld2018ContemporaryMetastases}. The success of managing multiple 
metastases interferes with different variables such as overall survival, 
quality of life, local tumor control, neurocognitive preservation, among 
several others. Depending on a patient\textquotesingle s overall health, the
stage and characteristics of the cancer, and other individual factors,
the medical team decides which treatment is most suitable for metastatic
cancer in the brain accounting for overall surviving rates as well as
treatment and follow-up care guidelines \cite{Garsa2021RadiationReview, 
Tsao2012RadiotherapeuticGuideline, Gondi2022RadiationGuideline, 
Mehta2005TheMetastases, Popat2021BrainEra, LeRhun2021EANO-ESMOTumours, 
Sahgal2020ACanada, Vogelbaum2022TreatmentGuideline}. These
guidelines are traced by several professional societies around the world
who gather experts from different fields to collaborate and share
knowledge, with the ultimate goal of improving cancer prevention,
diagnosis, and treatment. This review intends to (i) analyze the management 
of brain metastases considering individual patient cases; (ii) provide an 
overview of the guidelines from different societies related to the treatment 
and follow-up care associated with multiple brain metastases; (iii) assess 
state-of-the-art relating to the application of artificial intelligence 
algorithms capable of predicting the incidence of brain metastases after 
radiotherapy treatment.

\subsection{
\emph{Research Questions}} 
\paragraph{}
The major goal of this systematic review is to assess recent articles 
on management and follow-up recommendations for multiple brain 
metastases,
published between 2005 and May 2023. Therefore, the purpose of this
review is to respond to the following questions: 1) Are there specific
patient characteristics or tumor features that influence the choice of
treatment modality or affect treatment outcomes? 2) What are the
recommended follow-up protocols after treatment for multiple brain
metastases? 3) Are there artificial intelligence applications to 
predict recurrence of metastases during monitoring? 4) What gaps exist 
in the current literature and guidelines regarding the treatment and 
follow-up of patients with multiple brain metastases?

\subsection{
\emph{Search Strategy}} 
\paragraph{}
The methodology used for elaborating the current systematic review and
answer the previous questions was PRISMA. The search was performed in
multiple databases including PubMed, ScienceDirect and BioMed Central
with the search query `((``Management of multiple brain metastases'' OR
``Treatment of multiple brain metastases'' OR ``Multiple brain
metastases'' OR ``Multiple BMs'') AND (``Follow-up care of multiple
brain metastases'' OR ``Monitoring of multiple brain metastases'' OR ``Follow-up of multiple brain metastases after radiotherapy treatment'') AND
(``Guidelines for treatment of multiple brain metastases'' OR
``Recommendations for treatment of multiple brain metastases'' OR ``EANO
and ESMO guidelines for multiple brain metastases'' OR ``ASTRO and ASCO
guidelines for multiple brain metastases'' OR ``JASTRO guidelines for
multiple brain metastases'' OR ``DEGRO guidelines for multiple brain
metastases'' OR ``KSNO guidelines for multiple brain metastases'' OR
``Guidelines for follow-up care of multiple brain metastases'' OR
``Guidelines for monitoring multiple brain metastases'') AND
(``AI applications for predicting brain metastases'' OR ``AI to predict 
probability of brain metastases'' OR ``Artificial intelligence in forecasting 
recurrence of brain metastases'' OR ``Machine learning to predict incidence 
of brain metastases'' OR ``Predicting incidence of brain metastases'') AND
multiple brain metastases AND multiple BMs)' to find specific papers to
access the several topics previously mentioned. The articles covered in
this review were published after 2005 due to the ongoing upgrading and
transformation on the domain of accessing multiple brain metastases.
However, it should be noted that older papers are in minor number and
were mentioned in this review to enhance the advancements in the field.
\subparagraph{}
There were found a total of 138 records, of which 5 were removed from
being prior to 2005. From the remaining 133 records, 34 were then
excluded based on the titles and abstracts, that did not mention
treatments associated with multiple brain metastases. The resulting 99
papers were assessed and from these, 6 surveys were excluded due to lack
of substantial information relating to the research questions and other
4 papers did not present relevant results to be compared to others
works. This resulted in a total of 89 core papers, from which 13 are
other reviews regarding topics related to multiple BMs. The PRISMA  \cite{PRISMA}
diagram in Figure 1 provides a summary overview of the screening.
\begin{figure}[ht]
\centering
\includegraphics[width=12cm, height=10cm]{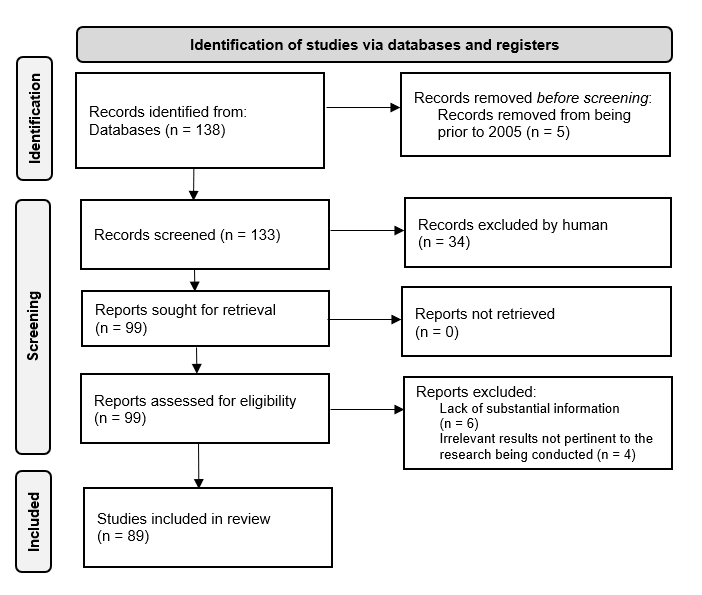}
\caption{Selection of the papers according to PRISMA diagram \cite{PRISMA}.}
\end{figure}

\paragraph{}
\paragraph{}
\paragraph{}

\subsection{
\emph{Manuscript Outline}} 
\paragraph{}
The current systematic review on the guidelines for treatment and
follow-up care of multiple brain metastases, includes an approach of the
general recommendations currently carried out by various societies
around the world on the management of metastases according to multiple
scenarios, as well as a general explanation of the treatments available
and comparisons between them, evaluating which one is better according
to a patient's circumstances. Furthermore, this review also explores the
usage of artificial intelligence in forecasting the incidence of
multiple metastases which would be an advancement in the strategy of 
treatment, helping medical professionals to explore different case scenarios 
and improve the outcomes for patients. This section presents the manuscript
outline, the search strategy, and the research questions. The rest of
this review is organized as follows.

\begin{itemize}
\item
  \emph{\textbf{Section 2}} covers the treatments for multiple brain
  metastases in great detail. A comparison of the treatments is made,
  and the best courses of action are determined based on the
  patient\textquotesingle s condition.
\item
  \emph{\textbf{Section 3}} discusses current techniques used in
  monitoring the disease after main treatment, including several medical
  image modalities.
\item
  \emph{\textbf{Section 4}} provides insight into the current guidelines
  by different societies around the world including EANO, ESMO, ASTRO,
  ASCO, SNO and KSNO on their management of multiple metastases and
  follow-up care considering various factors such as number, size and
  location of the tumors, and the patient's overall health among others.
\item
  \emph{\textbf{Section 5}} is the primary focus of this review since it
  provides a relatively fresh perspective on the potential application
  of artificial intelligence models to predict recurrence of brain
  metastases while the disease is being monitored, assisting healthcare
  professionals in early treatment planning.
\item
  Finally, \emph{\textbf{Section 6}}, concludes on the topics explored
  in the previous sections, emphasizing research options that should be
  pursued. It should be made clear that this is an overview focusing
  majorly on the treatment of numerous metastases. The motivation for
  this decision is the desire to enhance the application of artificial
  intelligence for the management of brain metastases, a poorly established
  area with significant untapped potential.
\end{itemize}

\subsection{
\emph{Acronyms and abbreviations}} 
The following Table 1 shows abbreviations that are used more than once
throughout the review. Other abbreviations are defined directly before
each table, when used only once.

\begin{longtable}{p{3cm}|p{7cm}}
\captionsetup{singlelinecheck=false, justification=raggedright}
\caption {Acronyms used throughout the review. \label{long}}\\
AHS &Adaptive Hybrid Surgery \\
\hline
AHSA & Adaptive Hybrid Surgery Analysis \\
\hline
ASCO & American Society of Clinical Oncology \\
\hline
ASTRO & American Society for Radiation Oncology \\
\hline
BMs & Brain Metastases \\
\hline
BTR & Brain Tumor Recurrence \\
\hline
CEST & Chemical Exchange Saturation Transfer \\
\hline
CITV & Cumulative Intracranial Tumor Volume \\
\hline
COVID-19 & Coronavirus Disease 2019 \\
\hline
CT & Computed Tomography \\
\hline
DVH & Dose Volume Histogram \\
\hline
DWI & Diffusion Weighted Imaging \\
\hline
EANO & European Association of Neuro-Oncology \\
\hline
EFS & Event-Free Survival \\
\hline
Elements MBM & Elements Multiple Brain Mets
(BrainLab\textsuperscript{®}) \\
\hline
ESMO & European Society for Medical Oncology \\
\hline
FSRS & Fractioned Stereotactic Radiosurgery \\
\hline
GI & Gradient Index \\
\hline
GTR & Gross Total Resection \\
\hline
GTV & Gross Tumor Volume \\
\hline
Gy & Gray \\
\hline
HA & Hyperarc \\
\hline
IMD & Intracranial Metastatic Disease \\
\hline
KPS & Karnofsky Performance Score \\
\hline
KSNO & Korean Society for Neuro-Oncology \\
\hline
LBRT & Local Brain Radiation Therapy \\
\hline
LGP & Leksell Gamma Plan \\
\hline
LMM & Leptomeningeal Metastases \\
\hline
LUAD & Lung Adenocarcinoma \\
\hline
MBM & Melanoma Brain Metastases \\
\hline
MRI & Magnetic Resonance Imaging \\
\hline
MRS & Magnetic Resonance Spectroscopy \\
\hline
MTR & Margin Total Resection \\
\hline
NSCLC & Non-Small Cell Lung Carcinoma \\
\hline
OS & Overall Survival \\
\hline
PET & Positron Emission Tomography \\
\hline
PI & Paddick Index \\
\hline
PTV & Planning Tumor Volume \\
\hline
PWI & Perfusion Weighted Imaging \\
\hline
RCTs & Randomized Controlled Trials \\
\hline
RD-WBRT & Reduced-dose Whole-brain Radiotherapy \\
\hline
SCLC & Small-cell Lung Cancer \\
\hline
SD-WBRT & Standard-dose Whole-brain Radiotherapy \\
\hline
SFRT & Stereotactic Fractioned Radiotherapy \\
\hline
SNO & Society for Neuro-Oncology \\
\hline
SRS & Stereotactic Radiosurgery \\
\hline
SRT & Stereotactic Radiotherapy \\
\hline
STR & Subtotal Resection \\
\hline
T1W & T1-Weighted \\
\hline
TTIP & Time To Intracranial Progression \\
\hline
VS & Vestibular Schwannoma \\
\hline
WBRT & Whole-Brain Radiation Therapy \\
\end{longtable}

\section{Treatment for Brain Metastases} \label{s:intro}
\paragraph{}
Treatments for brain metastases may ease symptoms, slow tumor growth,
and increase the patient's overall survival. Management options include
systemic therapies, surgery, stereotactic radiotherapy (SRT),
stereotactic radiosurgery (SRS), whole-brain radiation therapy (WBRT) 
or some combination of these. Nowadays treatments are adapted to 
individual cases and have the ultimate goal to reduce toxicity 
outcomes, while being the most effective, hence, promoting long term 
survival, as well as quality of life \cite{Lin2015TreatmentMetastases}. The medical team first 
gathers information including patient factors (such as age, sex, 
overall health, performance status), tumor factors (number and size of 
brain metastases, primary tumor type, extracranial disease activity), 
and available treatment options (such as access to neurosurgery or 
stereotactic radiosurgery) to accurately recommend the most suitable 
treatment \cite{Lin2015TreatmentMetastases, Tsao2012RadiotherapeuticGuideline, Fabi2011BrainCenter}. The following subsections 
(which are condensed in Figure 2) will go over several treatments.

\begin{figure}[ht]
\centering
\includegraphics[width=16cm, height=10cm]{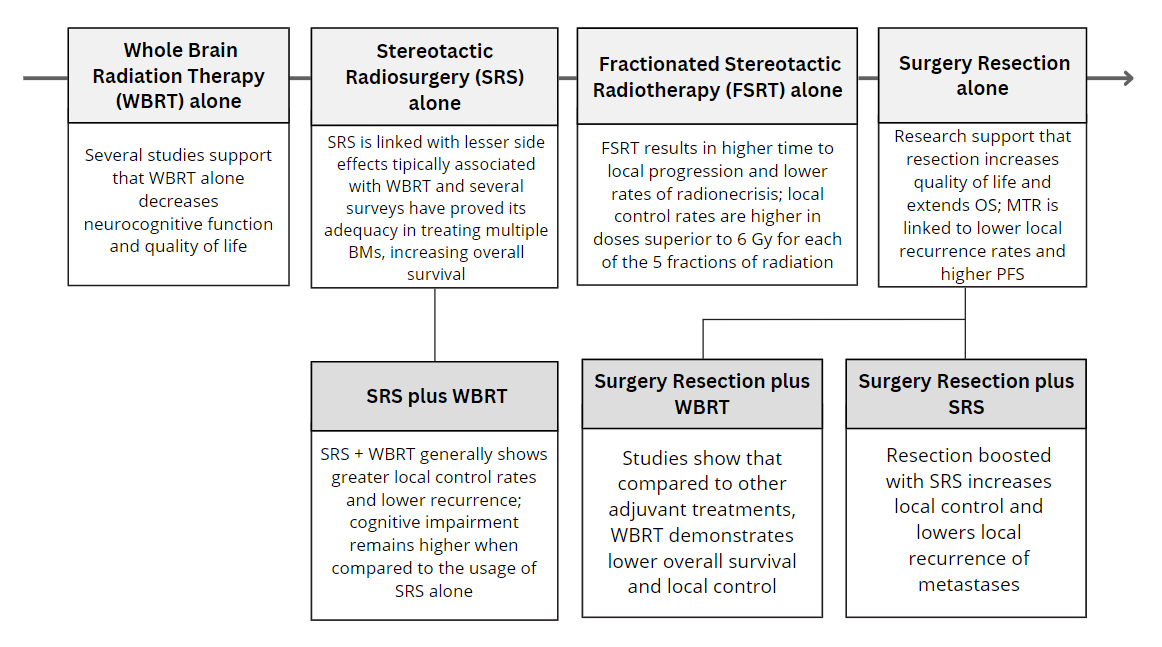}
\caption{Subsections covering different treatments of multiple brain metastases.}
\end{figure}

\subsection{
\emph{Whole Brain Radiation Therapy (WBRT) alone}}
\paragraph{}
Whole brain radiation therapy is a type of radiation treatment that
delivers radiation to the entire brain, rather than targeting a specific
area. Historically speaking, WBRT was considered the standard treatment
for patients with brain metastases, as it showed efficacy on metastatic
dissemination in the brain, provided good tumor control, and improved
symptomatology \cite{Lauko2020MedicalMetastases, Patel2014ManagementMetastasis, ChaoROENTGEN-RAYMETASTASES}. WBRT is typically advised for
patients with more than 3 brain metastases, with a dose of 30 Gy in 10
daily fractions. This because the treatment covers all brain tissue and
has the potential of irradiating yet non-visible lesions 
\cite{VanGrinsven2021TheMeta-Analysis, Lin2015TreatmentMetastases}. The use of 
WBRT, however, has decreased throughout the years
owing to the progress in radiation technology that allows targeted
radiation delivery, as well as the mounting apprehensions concerning the 
toxicity outcome effects linked with WBRT 
\cite{Lauko2020MedicalMetastases, Liu2019ManagementPresent, 
McTyre2013WholeMetastasis, Brown2018Whole-BrainRevolution}.
Studies (Table 2) have concluded that management of brain metastases
using WBRT alone can result in a decline of the patient's neurocognitive
capacity \cite{Yao2023TheRadiotherapy, Salzmann2022Long-termImmunotherapy}. Salzmann et al. \cite{Salzmann2022Long-termImmunotherapy} investigated a
cohort of 8 patients who had suffered from melanoma brain metastases
(MBM) and treated with WBRT to assess whether they had suffered
cognitive decline after treatment and analyze the long-term
neurocognitive effects of WBRT. Results showed that 5 patients
experienced little restriction on day-to-day life and the remaining 3
experienced cognitive decline associated with the area where the tumor
mass developed. The study was found to be limited for its small patient
sample, although it proved to be relevant for the topic's discussion.
More recently, Yao et al. \cite{Yao2023TheRadiotherapy} aimed to evaluate the 
correlations
between psychological distress, cognitive impairment, and quality of
life in patients with brain metastases after WBRT. The study enrolled 
71 patients with brain metastasis treated with WBRT, who were 
investigated with several scales before and after WBRT. Results showed 
that after receiving WBRT, the cognitive function and quality of life 
of patients decreased, while psychological distress increased.
\paragraph{}
The management techniques regarding multiple brain metastases are in
constant trial to expand the barriers of treatment approaches aiming
different scenarios (i.e., number and size of metastases, primary tumor
characteristics, among others). In a 2017 study, Khan et al. \cite{Khan2017ComparisonMeta-analysis}
compared the effectiveness of WBRT alone, SRS alone, and their
combination in the treatment of brain metastases based on randomized
controlled trial studies. Five studies (n=763) were included in this
analysis and the results showed that no significant survival benefit was
observed for any treatment approach, but local control was poorly
achieved when WBRT was applied alone, compared to its combination with
SRS. On the other hand, no difference in radiation-related toxicities
was found among the three approaches. Patil et al. \cite{Patil2017WholeMetastases} also
discussed the efficacy of WBRT plus SRS versus WBRT alone in the
treatment of brain metastases. By conducting a meta-analysis of two
studies, with a total of 358 participants, no difference in overall
survival (OS) between the two groups was found, similarly to the work
previously stated. However, participants with one brain metastasis had
significantly shorter median survival in the WBRT-alone group.
Participants in the combined therapy group had better local control,
performance status scores, and decreased steroid use compared to the
WBRT-alone group, as in the study by Khan et al. In 2022, Gaebe et al.
\cite{Gaebe2022StereotacticMeta-analysis} also compared the efficacy of SRS and WBRT in the treatment of
intracranial metastatic disease (IMD) in patients with small-cell lung
cancer (SCLC). A systematic review and meta-analysis of 31 studies were
conducted, and the primary outcome was overall survival. The results
showed that SRS had longer overall survival compared to WBRT alone or
with SRS boost, but not with WBRT plus SRS boost. Single-arm studies
reported a pooled median overall survival of 8.99 months. The study
suggests that survival outcomes are similar between SRS and WBRT in
treating IMD in SCLC patients, and future studies should focus on
differences in progression between the two treatments.
\paragraph{}
Whole brain radiation therapy is often used as adjunctive therapy or as
monotherapy for metastases that are not removable by surgery or
unresponsive to radiation. The treatment is often applied to patients
with multiple metastases ($>$3 metastases). Stereotactic
radiosurgery (SRS), on the other hand, has become a more widely used
treatment option for multiple brain metastases due to advancements in
technology, but upfront WBRT remains the standard approach for such
patients \cite{Lin2015TreatmentMetastases, Patel2014ManagementMetastasis, Brown2018Whole-BrainRevolution}.

\begin{longtable}{|p{2cm}|p{1cm}|p{4cm}|p{2cm}|p{5.4cm}|}
\captionsetup{singlelinecheck=false, justification=raggedright}
\caption{Studies and respective clinical outcomes regarding the application of WBRT alone as treatment of BMs.} \\
    \hline
    \multicolumn{1}{|c|}{\textbf{Study}} &
    \multicolumn{1}{c|}{\textbf{Year}} &
    \multicolumn{1}{c|}{\textbf{Purpose}} &
    \multicolumn{1}{c|}{\textbf{Arm}} &
    \multicolumn{1}{c|}{\textbf{Clinical Outcomes}} \\
    \hline
    \endfirsthead
    \hline \multicolumn{5}{r}{{(Continues on next page)}} \\ 
    \endfoot
    \hline
    \endlastfoot
    Khan et al. \cite{Khan2017ComparisonMeta-analysis} & 2017 & Compare the effectiveness of WBRT alone, SRS alone, and their combination in the treatment of BMs & SRS alone WBRT alone  SRS + WBRT & No significant survival benefit was observed for any treatment approach; local control was better in SRS+WBRT arm; no difference in radiation-related toxicities was found among the three approaches \\
    \hline
    Patil et al. \cite{Patil2017WholeMetastases} & 2017 & Compare the efficacy of WBRT + SRS versus WBRT alone in the treatment of BMs & WBRT alone SRS + WBRT & No difference in OS between the two groups; WBRT alone had shorter median survival in patients with 1 BM; WBRT+SRS had better local tumor control \\
    \hline
    Salzmann et al. \cite{Salzmann2022Long-termImmunotherapy} & 2022 & Assess long-term neurocognitive effects on patients diagnoses with MBM, treated with WBRT & WBRT & Three patients experienced cognitive decline associated with tumor location \\
    \hline
    Gaebe et al. \cite{Gaebe2022StereotacticMeta-analysis} & 2022 & Compare the efficacy of SRS + WBRT in the treatment of IMD associated with SCLC & SRS alone WBRT alone SRS + WBRT & Survival outcomes are similar in the SRS and WBRT arms \\
    \hline
    Yao et al. \cite{Yao2023TheRadiotherapy} & 2023 & Evaluate psychological distress, cognitive 
    impairment, and quality of life in patients with brain metastases after WBRT & WBRT & Cognitive function and quality of life, while 
    psychological distress increased \\
    \hline
\end{longtable}

\subsection{
\emph{Stereotactic Radiosurgery (SRS) alone}} 
\paragraph{}
Stereotactic radiosurgery (SRS) is a non-invasive technique for treating
both intracranial and extracranial lesions. Throughout history,
neurosurgeons and physicists developed various stereotactic devices,
imaging techniques, and radiation energy sources, however, it would only
be in 1951, that a Swedish neurosurgeon by the name of Lars Leksell
would come to introduce the concept of SRS. Over the course of the past
70 years, technological advancements greatly improved treatment
precision, patient comfort, and dose fractionation making SRS today a
highly versatile and cost-effective therapy \cite{Chen2005Stereotactic1, Kushnirsky2015TheRadiosurgery}.
\paragraph{}
Contrary to WBRT, SRS involves the precise focusing of radiation from
multiple angles and provides a confined area of high-dose radiation. By
using multiple, convergent beams of high energy, SRS decreases the dose of radiation reaching healthy tissue
and allows avoidance of radiation sensitive tissue like the optic nerve
\cite{Harris2022StereotacticRadiosurgery}. SRS has become a popular alternative to WBRT due to its
shorter and less invasive treatment course. It can be used alone for
patients with a limited number of BMs or in combination with WBRT,
systemic therapies, targeted therapies, and immunotherapies \cite{Badiyan2016StereotacticMetastases}.
Earlier research already favored the usage of SRS alone to preserve
neurocognitive function. Chang et al. \cite{Chang2009NeurocognitionTrial} compared the effects of
SRS alone to SRS plus WBRT on the learning and memory functions of
patients with 1-3 newly diagnosed BMs. The cohort included a total of 58
patients (n=30 in the SRS alone group, n=28 in the SRS plus WBRT group)
from which those who received SRS plus WBRT had significant chance of
learning and memory function decline at 4 months, compared to those
receiving SRS alone. Ultimately, the study showed that SRS alone entails
lesser toxicity effects usually associated with WBRT, making it a far
better choice as primary treatment for multiple metastases.
\paragraph{}
The improvement of radiation therapy and imaging technology, along with
the acknowledgement of the adverse effects of WBRT have sparked the
interest in using SRS as an alternative for patients with multiple BMs
\cite{Limon2017SingleMetastases}. Several clinical evidence (Table 3) could support such
statement \cite{Limon2017SingleMetastases, Mizuno2019ComparisonCancer, AmaanAliSurvivalLesions, Yamamoto2014StereotacticStudy}. A study published in 2014 by Yamamoto et
al. \cite{Yamamoto2014StereotacticStudy} aimed to determine whether SRS alone as the initial treatment for patients with 5-10 BMs was non-inferior to that for
patients with 2-4 BMs in terms of overall survival (OS). A total of
1,194 eligible patients were enrolled, and standard SRS procedures were
used in all patients. The study found that overall survival did not
differ significantly between patients with 2-4 metastases and those with
5-10. Hence, considering the minimal invasiveness of SRS and the minimal
outcome toxicity compared to WBRT, SRS alone proved to be a suitable
alternative for patients with up to 10 brain metastases. 
\paragraph{}
Following this investigation, Amaan Ali et al. \cite{AmaanAliSurvivalLesions} 
intended to understand whether the number of BMs alone had major impact on the choice between SRS or WBRT treatments. The investigation gathered a cohort of 5750 patients with BMs who underwent SRS. The researchers categorized the patients based on the number of BMs they had, with categories of 1, 2-4, 5-10, and $>$10 metastases. The median OS was compared between the
categories, and a multivariate analysis was performed to account for
other factors such as age, Karnofsky Performance Score (KPS), systemic
disease status, tumor histology, and cumulative intracranial tumor
volume (CITV). The results showed that patients with a single metastases
(1 BM) had superior median OS compared to those with 2-4 BMs (7.1 months
vs. 6.4 months, p=0.009); the OS in patients with 2-4 BMs was not
significantly different from those with 5-10 BMs (6.4 months v. 6.3
months, p=0.170). Patients with $>$10 BMs had lower median OS
than those with 2-10 BMs (5.5 months vs. 6.3 months, p=0.025). The
multivariate analysis showed that when comparing patients with 1 BM to
those with 2-10 BMs, or those with 10 BMs to those with $>$10
BMs, risk of death increased by 10\%. In other words, patients with a
higher number of BMs had a higher risk of death. On the other hand, when
the number of BMs was modeled as a continuous variable, the analysis
showed that there was a 5\% increase in the risk of death for every
increment of 5-6 BMs. This means that even small increases in the number
of BMs can have an impact on survival. Accordingly, the research
suggests that the number of BMs alone doesn't have primary impact on OS
and therefore should not be considered as a major factor when deciding
between SRS and WBRT. Other factors, such as total tumor volume,
prescribed dose, age, systemic disease status, among others, should also
be considered when making treatment decisions.
\paragraph{}
Treating each BM separately with radiation therapy can be time-consuming and may limit the feasibility of SRS for multiple BMs. Following this, a retrospective survey by Limon et al. \cite{Limon2017SingleMetastases} explored how the single-isocenter, multitarget (SIMT) for SRS planning and delivery would impact the treatment of BMs in order to identify those patients who might benefit the most from this procedure. The cohort included 59 patients with a median follow-up of 15.2 months and a median OS of 5.8 months. The study suggested that the number of metastases did not represent a major impact on OS. On the other hand, better OS was registered on planning tumor volume (PTV) $<$ 10 cc versus $\geq$ 10 cc (7.1 vs 4.2 months, respectively; $P = 0.0001$), which indicates the treatment works more effectively in smaller volumes. Another factor influencing OS was the dose administered to the PTV. Results showed that PTVs treated with $\geq$ 19 Gy were associated with increased overall survival. Conversely, patients receiving a dose of $>$12 Gy to $\geq$ 10 cc of normal brain had worse 
survival. Although it presented a limited follow-up period and a small cohort of patients, this study suggested that using SRS for the treatment of multiple BMs is effective and that outcomes such as OS might be less affected by the number of lesions and more by the tumor volume of those.
\paragraph{}
Another study conducted by Mizuno et al. \cite{Mizuno2019ComparisonCancer} 
suggests that SRS as an upfront treatment for 10-20 BMs could potentially delay the need for WBRT and therefore, its associated adverse events. This retrospective research intended to compare the efficacy and safety of SRS and WBRT in patients with advanced non‑small cell lung cancer (NSCLC) presenting
10‑20 BMs. The cohort included 44 patients with 10 to 20 brain
metastases from NSCLC who had been treated with SRS or WBRT as an
initial treatment. The findings indicated no significant difference in
OS between the two groups, however, the median time to intracranial
progression (TTIP) was significantly shorter in the SRS group than in
the WBRT group (7.1 months vs. 19.1 months, P=0.009), meaning the
disease progressed more rapidly in the SRS group. Neurological survival
did not differ between the two groups, and the type of initial treatment
(WBRT or SRS) was not found to be a significant prognostic factor in the
univariate and multivariate analyses. However, other factors such as
histology, performance status, subsequent molecular targeted drugs,
subsequent chemotherapy, and salvage treatment were identified as
independent prognostic factors. More recently, Shafie et al. 
\cite{ElShafie2020AMetastases}
also compared the effectiveness of SRS alone compared to WBRT alone for
treating multiple BMs. The research analyzed data from 128 patients over
a 5-year period, assessing various outcomes. Results revealed that
patients who received SRS had a median of 4 BMs, and the 1-year local
control of individual BM after SRS was roughly 92\%. Median OS was
significantly longer in the SRS arm, over 15 months, compared to the 8
months in the WBRT subgroup. The distant intracranial progression-free
survival (PFS) was shorter in the SRS subgroup (close to 9 months)
compared to the WBRT subgroup (about 22 months), indicating that
patients who receive WBRT might have lower hazard of distant
intracranial progression. The survey also found that synchronous BM
diagnosis, higher initial number of BMs and lung cancer histology, were
detrimental factors negatively affecting survival after SRS. Ultimately,
the study supported the use of SRS as a feasible and effective treatment
option for patients with multiple BMs. SRS alone was associated with
longer OS compared to WBRT.
\paragraph{}
More recent research by Ogawa et al. \cite{Ogawa2022ClinicalSurgery} assessed the 
effectiveness of fractionated stereotactic radiosurgery (FSRS) as a treatment for 
larger BMs ($\geq$ 20 mm). 105 patients with a total of 116 BMs 
were examined with a median maximum tumor diameter of 25 mm, and median 
prescribed dose of 35 Gy in 3 fractions. The findings revealed that 
FSRS achieved good local control with a local failure rate of 12.5\% 
and an intracranial failure rate within 1 year of nearly 57\%. Because 
of multiple and local recurrences, 21 and 4 patients had to be 
submitted to WBRT and surgery after FSRS, respectively. The study 
concluded that BMs greater than 20 mm present worthy local control when 
treated with fractioned SRS. Moreover, the eligibility of the survey 
proved that FSRS is a potential alternative to surgery since few 
patients were administered to surgery after FSRS treatment.
\paragraph{}
Recent research by Chea et al. \cite{Chea2021DosimetricCharacteristics} compared 
the efficacy of Linac-based mono-isocentric SRS with BrainLab\textsuperscript{®}
Elements Multiple Brain Mets (MBM) SRS to the Gamma Knife for the
treatment of patients with 3-9 BMs. The study involved 20 patients who
were previously treated with Gamma Knife SRS. The patients had 3 to 9
BMs from different primary malignant tumors. Totally there were 95
metastases with a major axis diameter ranging from 0.3 to 4 cm and
volume ranging from 0.02 to 9.61 cc. Various statistical methods were
used to evaluate the dosimetric impact of target volume geometric
characteristics, including the Paddick Index (PI), Gradient Index (GI),
dose fall-off, volume of healthy brain receiving more than 12 Gy, and
dose volume histogram (DVH). Results showed that both approaches have
similar plan qualities, with MBM having somewhat better selectivity for
smaller lesions but Leksell Gamma Plan (LGP) having slightly better healthy
tissue sparing at the cost of a significantly longer irradiation period.
For larger volumes, MBM strategies must take a higher healthy tissue
exposure into account. Several research supports the usage of Elements
MBM in generating high quality plans for treatment of multiple brain
metastases. Raza et al. \cite{Raza2022Single-isocenterSystems} concluded that 
MBM had the same capacity to generate high quality plans in patients presenting 
up to 25 BMs, as the Varian Hyperarc (HA) system. Additionally, Cui et al. 
\cite{Cui2022EvaluationIsocenter} established that compared to knowledge-based 
planning (KBP), the MBM plans had a higher capacity to spare normal brain tissues 
in terms of total volumes receiving $\geq$ 5 Gy.
\paragraph{}
When managing multiple BMs, plenty of studies support the efficacy of
treatment with SRS as a far better choice than WBRT, regarding the side
effects. However, the number of brain metastases alone is not sufficient
factor to determine the best treatment approach and medical
professionals should account for several other aspects such as age,
histology, performance status, tumor volume and prescribed dose.

\begin{longtable}{|p{2cm}|p{1cm}|p{4cm}|p{2cm}|p{5.4cm}|}
\captionsetup{singlelinecheck=false, justification=raggedright}
\caption{Studies and respective clinical outcomes regarding the application of 
SRS alone as treatment of BMs.}\\
    \hline
    \multicolumn{1}{|c|}{\textbf{Study}} &
    \multicolumn{1}{c|}{\textbf{Year}} &
    \multicolumn{1}{c|}{\textbf{Purpose}} &
    \multicolumn{1}{c|}{\textbf{Arm}} &
    \multicolumn{1}{c|}{\textbf{Clinical Outcomes}} \\
    \hline
    \endfirsthead
    \hline \multicolumn{5}{r}{{(Continues on next page)}} \\ 
    \endfoot
    \hline
    \endlastfoot
    Chang et al. \cite{Chang2009NeurocognitionTrial} & 2009 & Compare the effects 
    of SRS alone and SRS + WBRT on the cognitive functions of patients with 1-3 
    newly diagnosed BMs & SRS alone SRS+WBRT & SRS alone entailed lesser toxicity 
    effects usually associated with WBRT \\
    \hline
    Yamamoto et al. \cite{Yamamoto2014StereotacticStudy} & 2014 & Compare OS in 
    patients receiving SRS alone as the initial treatment of 5-10 BMs and 2-4 BMs 
    & SRS alone & OS did not differ significantly between patients with 2-4 BMs 
    and 5-10 BMs; SRS alone proved to be a suitable alternative for patients with 
    up to 10 BMs  \\
    \hline
    Amaan Ali et al. \cite{AmaanAliSurvivalLesions} & 2017 & Understand whether 
    the number of BMs alone had a major impact on the choice between SRS or WBRT 
    & SRS alone WBRT alone SRS + WBRT & Patients with 1 BM had superior OS 
    compared to those with 2-4 BMs (7.1 months vs. 6.4 months); OS in patients 
    with 2-4 BMs was not significantly different from those with 5-10 BMs (6.4 
    months vs. 6.3 months); patients with $>$10 BMs had lower median OS \\
    \hline
    Limon et al. \cite{Limon2017SingleMetastases} & 2017 & Understand how the 
    single-isocenter, multitarget (SIMT) for SRS planning and delivery would 
    impact the treatment of BMs & SIMT SRS & Number of metastases did not 
    represent a major impact on OS; Better OS was registered on PTV $<$10 cc 
    versus $\geq$10 cc (7.1 vs. 4.2 months, respectively); PTVs treated with 
    $\geq$19 Gy were associated with increased OS \\
    \hline
    Mizuno et al. \cite{Mizuno2019ComparisonCancer} & 2019 & Compare the efficacy 
    and safety of SRS and WBRT in patients with advanced non‑small cell lung 
    cancer (NSCLC) presenting 10‑20 BMs & SRS alone WBRT alone & Median TTIP was 
    significantly shorter in the SRS arm compared to the WBRT arm (7.1 months vs. 
    19.1 months); The type of initial treatment was not found to be a significant 
    prognostic factor \\
    \hline
    Shafie et al. \cite{ElShafie2020AMetastases} & 2020 & Compare the 
    effectiveness of SRS alone vs WBRT alone for treating multiple BMs & SRS 
    alone  WBRT alone & Median OS was longer in the SRS arm, more than 15 
    months, compared to 8 months in the WBRT arm \\
    \hline
    Ogawa et al. \cite{Ogawa2022ClinicalSurgery} & 2022 & Assess the 
    effectiveness of fractionated SRS as a treatment for larger BMs ($\geq$20 mm) 
    & FSRS (35 Gy in 3 fractions) & FSRS achieved good local control with a local 
    failure rate of 12.5\% and an intracranial failure rate within 1 year of 
    nearly 57\% \\
    \hline
\end{longtable}

\subsection{
\emph{Fractionated Stereotactic Radiotherapy (SFRT)}}
\paragraph{}
Fractionated stereotactic radiotherapy (FSRT) is a specialized technique
in radiation therapy that combines the precision of stereotactic
radiosurgery (SRS) with the delivery of radiation in multiple fractions.
For patients with a small number of brain metastases, SRS is an enticing
alternative. However, the efficacy and safety of SRS might be
jeopardized when treating large lesions or those close to healthy
tissue. In these situations, FSRT is frequently employed in an effort to
raise the ratio between the desired therapeutic effects and the
potential side effects caused by radiation \cite{Baliga2017FractionatedModelling, 
Lupattelli2022StereotacticOligometastases}. Several clinical trials (Table 4) 
have proven the efficacy of FSRT, namely five-fraction SRT, as an advantageous 
alternative for managing multiple lesions, especially regarding LC and 
radionecrosis rates.
\paragraph{}
A study by Putz et al. \cite{Putz2020FSRTStudy} compared the effectiveness and 
safety of FSRT and SRS for the treatment of BMs. The survey included a cohort of
120 patients (n=98 treated with FSRT and n=92 treated with SRS). Results
showed that the biologically effective dose (BED) for metastases was
related to local control, while the BED for normal brain tissue was
associated with radionecrosis. The median time to local progression was
nearly 23 months in the FSRT group compared to the 14.5 months in the
SRS arm. There was a significant difference in overall rate of
radionecrosis at 12 months with 3.4\% vs. 14.8\% for FSRT and SRS,
respectively. The incidence of severe radionecrosis requiring resection
was also significantly lower in the FSRT group. The study proved the
efficacy of FSRT in treating both large and small BMs, highlighting its
potential in local control and reduced hazard of radionecrosis.
\paragraph{}
A retrospective survey by Piras et al. \cite{Piras2022Five-FractionSchedules} 
investigated the optimal dose schedule for five-fraction SRT in the treatment of 
BMs. The analysis was conducted on 41 patients treated with different
five-fraction SRT dose schedules, administered over 5 consecutive days,
with prescribed doses ranging from 30 to 40 Gy and covering at least
98\% of the gross tumor volume (GTV). Magnetic resonance imaging (MRI)
was performed at 3, 6, and 9 months to evaluate LC and clinical outcomes
and data related to toxicity was also acquired. The evidence suggested
that higher LC rates were found when dose schedules exceeded 6 Gy per
fraction and that the toxicity rates were generally mild. Based on these
findings, the study concludes that the management of BMs with
five-fraction SRT is feasible. The analysis also suggests that a total
dose of less than 30 Gy in 5 fractions should be avoided due to the
expectation of lower LC. Another research by 
Layer et al. \cite{Layer2023Five-FractionAnalysis} also evaluated the safety of five-fraction SRT for the treatment of BMs as a definitive or adjuvant treatment. The study analyzed data from 36 patients who received FSRT (5 fractions of 7 Gy each) to BMs or cavities
after resection. Findings showed that the overall RN rate was more than
14\%, and the median time to RN occurrence was almost 13 months. Factors associated with RN occurrence were immunotherapy, young age ($\leq$45 years), and larger planning target volume (PTV). The cumulative 1-year LC rate
was over 83\%, indicating effective tumor control. The estimated median
LPF survival was roughly 19 months. Ultimately, the study concluded that
FSRT is a feasible and safe treatment approach for brain metastases. It
demonstrated acceptable local control rates and comparable rates of
radiation necrosis in both adjuvant and definitive radiotherapy
settings. Similar to the previous paper by Piras et al. this research
also supports the efficiency of five-fraction SRT as a great treatment
option for patients with BMs.
\paragraph{}
Following this study, Ding et al. \cite{Ding2023AnMetastases} intended to evaluate the feasibility and potential benefits of online adaptive MR-guided FSRT for patients with BMs. This survey included 28 patients diagnosed with BMs,
treated with FSRT (30 Gy in 5 fractions) using a 1.5 T MR-Linac. Daily
MRI scans were used, and customized treatment plans based on the
contours identified from the MR images were developed. For all lesions
assessed, the results demonstrated a significant decrease in tumor
volume during FSRT in comparison to the pre-treatment simulation. The
inter-fractional alterations in lesions with perilesional edema, which
accounted for more than 53\% of the lesions, were substantially
different from those in lesions without it. Compared to patients with a
single metastasis, patients with multiple metastases experienced more
inter-fractional tumor alterations, such as tumor volume reduction and
anatomical shift. In terms of treatment planning, 19\% of the fractions
in the non-adaptive plans showed insufficient PTV coverage. The study
concluded that significant inter-fractional tumor changes can occur
during FSRT, and that the daily MR-guided re-optimization of treatment
plans provided dosimetric benefits, particularly for patients with
perilesional edema or multiple lesions. This highlights the potential
advantages of online adaptive MR-guided FSRT in optimizing treatment
delivery and maintaining target coverage during the course of treatment.

\begin{longtable}{|p{2cm}|p{1cm}|p{4cm}|p{2cm}|p{5.4cm}|}
\captionsetup{singlelinecheck=false, justification=raggedright}
\caption{Studies and respective clinical outcomes regarding the application of fractionated stereotactic radiotherapy (FSRT) alone as treatment of BMs.} \\
    \hline
    \multicolumn{1}{|c|}{\textbf{Study}} &
    \multicolumn{1}{c|}{\textbf{Year}} &
    \multicolumn{1}{c|}{\textbf{Purpose}} &
    \multicolumn{1}{c|}{\textbf{Arm}} &
    \multicolumn{1}{c|}{\textbf{Clinical Outcomes}} \\
    \hline
    \endfirsthead
    \hline \multicolumn{5}{r}{{(Continues on next page)}} \\ 
    \endfoot
    \hline
    \endlastfoot
    Putz et al. \cite{Putz2020FSRTStudy} & 2020 & Compare the efficacy of FSRT and SRS in the treatment of BMs & FSRT alone SRS alone & Median time to local progression was 22.9 months in the FSRT group and 14.5 months in the SRS arm; there was significant difference in overall rate of radionecrosis at 12 months (3.4\% for FSRT vs. 14.8\% for SRS); the incidence of severe radionecrosis was significantly lower in the FSRT group \\
    \hline
    Piras et al. \cite{Piras2022Five-FractionSchedules} & 2022 & Investigate the optimal dose schedule for five-SRT in the treatment of BMs & FFSRT alone & Higher LC rates for dose schedules superior to 6 Gy per fraction; toxicity rates were not found to be higher than Grade 1 (mild) \\
    \hline
    Layer et al. \cite{Layer2023Five-FractionAnalysis} & 2023 & Determine the safety of five-fraction SRT for BMs, either as a unique or adjuvant treatment & FFSRT alone Resection + FFSRT & Overall RN rate was 14.3\%; median time to RN was roughly 12.9 months; RN occurrence was associated with immunotherapy, young age ($\leq$45 years), and large PTV;  estimated cumulative 1-year LC rate was 83.1\%; the median local PFS was 18.8 months; estimated median OS was 11 months \\
    \hline
    Ding et al. \cite{Ding2023AnMetastases} & 2023 & Assess the potential benefits of online adaptive MR-guided FSRT in patients with BMs & FSRT alone & Significant tumor volume reduction was found during FSRT compared to initial fraction; patients with multiple lesions had more significant inter-fractional tumor changes than those with single lesion \\ \hline
\end{longtable}

\subsection{
\emph{Surgery Resection}}
\paragraph{}
Surgical resection is one of the treatment options for brain metastases
and involves the removal of the tumor from the brain by surgical
procedure, without damaging regions responsible for critical brain
functions, surrounding the lesions. The decision to perform surgery
depends on several factors, including the number, size, location of the
tumors, the patient\textquotesingle s overall health, and the presence
of symptoms such as neurological deficits \cite{BrainReview, Patel2014ManagementMetastasis}. Although
there aren't many prospective studies confirming the effectiveness of
surgery resection in the treatment of multiple BMs, there is growing
evidence that it can improve functional outcomes and increase the
possibility of receiving additional therapies that are important for
overall cancer prognosis, such as SRS or systemic therapies. Although
surgical resection is generally recommended for patients with a single
BM, studies suggest that this procedure can also be effective when
applied to multiple metastases. Moreover, surgical resection can improve
clinical status and overall survival rates, improve quality of life and
patient's neurocognitive ability (Table 5) \cite{Patel2014ManagementMetastasis, Ene2022SurgicalNuances, Schodel2020SurgicalTreatment, Junger2021ResectionMetastases, Winther2022SurgeryResection}.
\paragraph{}
A retrospective study by Schödel et al. \cite{Schodel2020SurgicalTreatment} 
investigated the relationship between surgical resection of BMs and subsequent 
treatment and clinical outcomes. The research concluded that surgical resection
could improve clinical status and enhance the probability of receiving a
second treatment, leading to an increased overall survival. The authors
analyzed a large cohort of 750 patients who underwent resection of
symptomatic BMs. Results showed that surgical resection significantly
improved patients\textquotesingle{} functional status, with a median
Karnofsky Performance Score of 80 before surgery, increasing to 90 after
surgery. Hence, after resection patients presented a higher level of
function and performance of daily activities independently. Also,
patients who received postoperative local radiotherapy and systemic
treatment had significantly longer survival. Another research carried
out by Jünger et al. \cite{Junger2021ResectionMetastases} also suggested that 
resection of symptomatic BMs could be indicated to patients with multiple BMs to 
ease their neurological symptoms and facilitate further consecutive
treatments. The study gathered a cohort of 216 patients (n=129 were
diagnosed with a single BM; n=64 had 2-3 BMs; n=23 had more than 3 BMs)
who underwent surgical resection and collected demographic, clinical,
and tumor-associated parameters. Similar to Schödel et al., they found
that surgical resection of symptomatic BMs significantly improved the
KPS and enabled adjuvant radiotherapy and systemic treatment for up to
90\% and 55\% of patients, respectively. Among other results, the number
of BMs did not influence the local control rates and multiple prognostic
factors such as age, preoperative and postoperative KPS, presence of
extracranial metastases and consecutive radiation therapy and systemic
treatment influenced overall survival. In a more recent study, Winther
et al. \cite{Winther2022SurgeryResection} concluded that overall survival rates 
were improved with total resection of GTV compared to subtotal resection. The 
researchers reviewed 373 adult patients who underwent surgery for a single BM and
found that gross total resection (complete removal of visible tumor) was
associated with longer overall survival (13.0 months) compared to
subtotal resection (8.0 months). In addition to the previous studies,
these findings suggested that the extent of surgical resection also
takes an important factor in the management of BMs allowing to improve
overall survival.
\paragraph{}
Patients with BMs are becoming more prevalent as a result of improved
treatment options, growing medical standards, and ongoing development.
Although surgical resection of original BMs is well defined and
supported by numerous research, the use of resection for recurring BMs
is still poorly understood \cite{Ene2022SurgicalNuances, Heler2022RecurrentSetting}. Following the previous
study by Winther et al., Gong et al. \cite{Gong2022RecurrenceAdenocarcinoma} further intended to
investigate whether margin total resection (MTR) could reduce the local
recurrence rate of BMs from lung adenocarcinoma compared to gross total
resection (GTR). This retrospective research involved 48 patients
diagnosed with BMs from lung adenocarcinoma (LUAD). The variable under
study was the local tumor control from GTR and MTR. The MTR group
consisted of patients who had undergone GTR and had the tumor periphery
resected for more 5 mm. Results showed that the local recurrence rates
were significantly lower in the MTR group, for 6 months after surgery
the local recurrence rate was roughly 14\% in the MTR group compared to
around 42\% in the GTR group. The median progression-free survival rate
after surgery was twice as high in the MTR group compared to the GTR
group, 14.0 months, and 7.0 months, respectively. In conclusion,
although resection of the gross tumor volume might improve overall
survival, expanded peripheral resection of 5 mm around the BMs can
significantly reduce the local recurrence rate and prolong
progression-free survival time.
\paragraph{}
A retrospective study performed by Heßler et al. \cite{Heler2022RecurrentSetting} 
evaluated the effectiveness of surgical resection for recurrent symptomatic BMs. 
The research gathered a total of 107 patients with multiple occurrences of
primary tumors (non‑small cell lung cancer (NSCLC), breast, melanoma,
gastro‑intestinal, among others) who were also previously individually
treated. The variables at study were intracranial event-free survival
(EFS) and overall survival (OS), and results showed that the median
postoperative EFS and OS were roughly 7 and 11 months, respectively. The
median pre‑operative KPS was 70\%, improving to 80\% after surgery,
meaning that resection increased patient\textquotesingle s functional
status and independence to perform activities of daily life. However,
complication rate was greater than 26\%. Furthermore, the patient's
clinical status proved to be the only factor to remain independent from
survival analysis. The study suggested that although surgical resection
of recurrent BMs might improve the clinical status and OS, it is
associated with a high complication rate. Therefore, it is crucial to
carefully select patients before deciding on surgical resection for
recurrent BMs.

\begin{longtable}{|p{2cm}|p{1cm}|p{4cm}|p{2cm}|p{5.5cm}|}
\captionsetup{singlelinecheck=false, justification=raggedright}
\caption{Studies and respective clinical outcomes regarding the application of Surgery Resection alone as treatment of BMs.} \\
    \hline
    \multicolumn{1}{|c|}{\textbf{Study}} &
    \multicolumn{1}{c|}{\textbf{Year}} &
    \multicolumn{1}{c|}{\textbf{Purpose}} &
    \multicolumn{1}{c|}{\textbf{Arm}} &
    \multicolumn{1}{c|}{\textbf{Clinical Outcomes}} \\
    \hline
    \endfirsthead
    \multicolumn{5}{r}{{(Continues on next page)}} \\ 
    \endfoot
    \hline
    \endlastfoot
    Schödel et al. \cite{Schodel2020SurgicalTreatment} & 2020 & Understand the relationship between surgical resection of BMs and subsequent treatment, and clinical outcomes & Resection alone & After resection patients presented higher level of function; patients who received postoperative treatment had significantly longer survival \\
    \hline
    Jünger et al. \cite{Junger2021ResectionMetastases} & 2021 & Evaluate the role of surgical resection of BMs in patients with NSCLC, regardless of the number of lesions & Resection alone & Significant improvement in KPS after resection; BM count did not significantly influence local control rates; the mean OS after surgery was 12.7 months \\
    \hline
    Winther et al. \cite{Winther2022SurgeryResection} & 2022 & Investigate the association between OS and residual tumor after surgery for single BM & Resection alone & Median OS was 8.0 months for patients with subtotal resection and 13.0 months for patients with gross total resection \\
    \hline
    Gong et al. \cite{Gong2022RecurrenceAdenocarcinoma} & 2022 & Investigate whether MTR can reduce the local recurrence rate of BM from lung adenocarcinoma compared with GTR & Marginal Tumor Resection alone Gross Tumor Resection alone & Local recurrence rates 6 months after surgery: GTR group was 42.3\%, MTR group was 13.6\%. Local recurrence rates 12 months after surgery: GTR group was 57.7\%, MTR group was 22.7\%. Median progression-free survival time after surgery: GTR group was 7.0 months, MTR group was 14.0 months \\
    \hline
    Heßler et al. \cite{Heler2022RecurrentSetting} & 2022 & Evaluate the efficacy of surgery for pretreated, recurrent, and symptomatic BMs & Resection alone & Median KPS was 80\% after surgery; median postoperative EFS and OS were 7.1 and 11.1 months, respectively; clinical status remained the only independent factor for survival in multivariate analysis \\
    \hline
\end{longtable}

\subsection{
\emph{Stereotactic Radiosurgery (SRS) plus WBRT}}
\paragraph{}
Treating brain metastases with WBRT alone has been the standard approach
however, the toxicity outcomes have led to a wide appliance of SRS to
manage multiple metastases \cite{Limon2017SingleMetastases, Mizuno2019ComparisonCancer, AmaanAliSurvivalLesions, Yamamoto2014StereotacticStudy, Nakano2022Reduced-doseControl}. Nonetheless,
when WBRT is not used, there is a greater chance of brain tumor
recurrence (BTR), which could potentially negatively impact the overall
survival of certain patients \cite{Nakano2022Reduced-doseControl}. Results from a 
study led by Stafinski et al. \cite{Stafinski2006EffectivenessMeta-analysis} 
showed that not only did combining SRS with WBRT improve OS in patients with a 
single BM, but also improved local tumor control and functional independence. The 
research intended to assess the effectiveness of SRS alone or in combination with 
WBRT, compared to surgery and/or WBRT in prolonging survival and improving
quality of life and functional status of patients with BMs. Patients who
presented multiple BMs, did not show any difference in OS when treated
with WBRT plus SRS, compared to treatment with WBRT alone. However, in
patients who had a single metastasis, the same combination therapy
resulted in a significant improvement in survival rates. In terms of
controlling the tumor locally over a 24-month period, the rates were
significantly higher in the group of patients receiving WBRT plus SRS,
regardless of the number of metastases. A study by Chang et al. \cite{Chang2009NeurocognitionTrial}
mentioned above, also compared the effectiveness of SRS plus WBRT
compared to SRS alone. Although in terms of cognitive decline, SRS alone
proved to have far better response, patients who had received SRS plus
WBRT presented a higher rate of freedom from CNS recurrence at 1 year
(73\% of patients), compared to those who had received SRS treatment
alone (27\% of patients). This means that combining SRS with WBRT
resulted in absence of cancer recurrence in the brain or spinal cord
after treatment. When comparing SRS alone, WBRT alone and SRS plus WBRT,
any of the treatments offer significant survival benefit. Nonetheless,
SRS combined with WBRT has proven to be effective when it comes to
control of local tumor volume and extended brain tumor recurrence (BTR)
free time, period during which a patient does not experience the
recurrence of brain tumors after being treated 
\cite{Stafinski2006EffectivenessMeta-analysis, Khan2017ComparisonMeta-analysis, 
Qie2018StereotacticTrials}.
\paragraph{}
Another research conducted by Hasan et al. \cite{Hasan2014TheMetastases} 
determined the effectiveness of adding WBRT following SRS for treating brain
metastases. The study analyzed three randomized controlled trials,
involving 389 patients with 1 to 4 brain metastases, and five
retrospective studies, reporting on outcomes such as survival, control,
salvage therapy, and quality of life measures. The study found that
adjuvant WBRT following SRS was more effective at tumor local control
(1-year local control was roughly 90\% for SRS + WBRT and little above
70\% for SRS alone) and distant recurrence (mean crude distant
recurrence rate for SRS + WBRT was near 40\% and 54\% for SRS alone)
than SRS alone. However, the study also found that SRS plus adjuvant
WBRT did not benefit in terms of overall survival (mean 1-year survival
was little above 33\% for SRS + WBRT and roughly 39\% for SRS alone) or
symptomology. The addition of WBRT was linked to a lower quality of
life, and its known link to cognitive decline and neurotoxicity should
be evaluated against the advantage of local control.
\paragraph{}
So far, the analysis of the three works reinforces the debated character
of combining these treatments to manage brain metastases. The works
suggest contrary findings in terms of the influence of SRS plus WBRT in
terms of overall survival and preservation of neurocognitive ability
depending on the number of metastases to be cured. Moreover, all the
data indicated that in terms of recurrence and capacity of local control
this combination seems to have greater effect than implementing SRS
alone. In a later retrospective investigation by Brown et al. 
\cite{Brown2016EffectTrial} concluded that for patients with 1 to 3 BMs, the use 
of SRS alone, compared with SRS combined with WBRT, resulted in less cognitive
deterioration. The study also suggested that there was no difference in
terms of overall survival between the two groups. The work's primary
quest was to answer the debating question of whether there could be less
cognitive deterioration at SRS alone compared to SRS plus WBRT. The
research involved 213 participants who randomly received either SRS
alone or SRS plus WBRT. Findings indicated that there was less cognitive
deterioration at 3 months after SRS alone (roughly 64\% of patients)
than when combined with WBRT (nearly 92\% of patients) and quality of
life was also higher at 3 months with SRS alone.
\paragraph{}
Many studies regarding treatment with SRS plus WBRT present inconclusive
results regarding overall survival rates. Following this, Khan et al.
\cite{Khan2019WholeFactors} assessed whether the improved local control achieved 
with WBRT plus SRS leads to any survival benefit in patients with BMs and
favorable prognostic factors. Five studies (n=2728) were identified, and
the primary outcome was overall survival. It became apparent that WBRT
plus SRS improved survival in brain metastatic cancer patients with
better prognostic factors, particularly when compared to WBRT alone.
Nonetheless, the survival advantage over SRS was only limited to
non-small cell lung cancer primary tumor histology. This way, the
research suggests that WBRT combined with SRS might indeed improve
overall survival, considering selected patients with favorable
prognostic factors. As seen previously, WBRT has an impact in tumor
volume control, despite its toxicity side effects on cognitive function.
A more recent investigation undertaken by 
Nakano et al. \cite{Nakano2022Reduced-doseControl}, intended to assess whether combining reduced-dose whole-brain
radiotherapy (RD-WBRT) with SRS, could potentially minimize the risk of
cognitive decline, without compromising the brain tumor control for
patients with 1-4 BMs. The standard-dose WBRT (SD-WBRT) consists of 30
Gy in 10 fractions, whereas RD-WBRT consists of 25 Gy in 10 fractions).
The study enrolled 40 patients from seven different institutions (28
patients with primary tumor in the lung; 20 patients with single BM).
Results showed that the median OS time was 19 months; more than 75\% of
the patients didn't experience the recurrence of brain tumors in distant
locations at 6 months after treatment; 23\% of patients experienced the
recurrence of brain tumor at any location at 6 months after treatment,
considering the possibility of death; and roughly 49\% of patients
experienced a decline in cognitive function that persisted over time.
The work showed that RD-WBRT combined with SRS may reduce the risk of
cognitive decline when compared to standard WBRT.
\paragraph{}
The issue with combining SRS with WBRT to treat brain metastases
prevails, as incoherence whether this treatment option influences
overall survival, neurocognitive decline, local control, and metastases
recurrence persist. Moreover, results vary attending the number of BMs
treated. Several studies (Table 6) refer WBRT combined with SRS, to have
great effect on patients' local tumor control over time and preventing
metastases recurrence, however because of the different results
concerning neurotoxicity and influence on survival, this topic remains
unclear and in need of further research.

\begin{longtable}{|p{2cm}|p{1cm}|p{4cm}|p{2cm}|p{5.4cm}|}
\captionsetup{singlelinecheck=false, justification=raggedright}
\caption{Studies and respective clinical outcomes regarding the application of Stereotactic Radiosurgery (SRS) plus WBRT as treatment of BMs.} \\
    \hline
    \multicolumn{1}{|c|}{\textbf{Study}} &
    \multicolumn{1}{c|}{\textbf{Year}} &
    \multicolumn{1}{c|}{\textbf{Purpose}} &
    \multicolumn{1}{c|}{\textbf{Arm}} &
    \multicolumn{1}{c|}{\textbf{Clinical Outcomes}} \\
    \hline
    \endfirsthead
    \hline \multicolumn{5}{r}{{(Continues on next page)}} \\ 
    \endfoot
    \hline
    \endlastfoot
    Stafinski et al. \cite{Stafinski2006EffectivenessMeta-analysis} & 2006 & Assess the effectiveness of SRS alone or SRS+WBRT, compared to surgery and/or WBRT in prolonging survival and improving quality of life and functional status of patients with BMs & SRS alone SRS+WBRT Resection alone Resection + WBRT & In patients with a single BM SRS+WBRT improved OS, local tumor control and functional independence \\
    \hline
    Chang et al. \cite{Chang2009NeurocognitionTrial} & 2009 & Compare the effects of SRS alone and SRS + WBRT on the cognitive functions of patients with 1-3 newly diagnosed BMs	& SRS alone SRS+WBRT & SRS alone entailed lesser toxicity effects usually associated with WBRT \\
    \hline
    Hasan et al. \cite{Hasan2014TheMetastases} & 2014 & Determine the effectiveness of SRS+WBRT for treating 1 to 4 BMs, reporting on outcomes such as survival, control, salvage therapy, and quality of life measures & SRS alone SRS+WBRT & At 1-year local control was roughly 90\% for SRS + WBRT and little above 70\% for SRS alone; mean crude distant recurrence rate for SRS + WBRT was near 40\% and 54\% for SRS alone; SRS+WBRT did not benefit in terms of OS or symptomology \\
    \hline
    Brown et al. \cite{Brown2016EffectTrial} & 2016 & Answer the debating question of whether there could be less cognitive deterioration at SRS alone compared to SRS plus WBRT & SRS alone SRS+WBRT & There was less cognitive deterioration at 3 months after SRS alone (roughly 64\% of patients) than when combined with WBRT (nearly 92\% of patients); quality of life was higher at 3 months with SRS alone \\
    \hline
    Khan et al. \cite{Khan2019WholeFactors} & 2019 & Assess whether improved local control achieved with WBRT plus SRS leads to any survival benefit in patients with BMs and favorable prognostic factors & SRS+WBRT & WBRT+SRS might improve OS, considering selected patients with favorable prognostic factors \\
    \hline
    Nakano et al. \cite{Nakano2022Reduced-doseControl} & 2022 & Assess whether combining reduced-dose whole-brain radiotherapy (RD-WBRT) with SRS could potentially minimize the risk of cognitive decline, without compromising the brain tumor control for patients with 1–4 BMs & RD-WBRT (25 Gy in 10 fractions) + SRS & Median OS time was 19 months; more than 75\% of the patients didn’t experience the recurrence of brain tumors in distant locations at 6 months after treatment; roughly 49\% of patients experienced a decline in cognitive function that persisted over time \\
    \hline
\end{longtable}

\subsection{
\emph{Surgery Resection plus WBRT}}
\paragraph{}
SRS is frequently preferred to WBRT since neurocognition and quality of
life are increasingly being used as key therapy objectives in patients
with brain metastases, even though WBRT affords significant intracranial
disease control by targeting the entire brain 
\cite{Patel2014ManagementMetastasis, Vlachos2023StereotacticMeta-analysis, 
Nieder2023BrainTherapy}. Moreover, while SRS might increase the hazard of distant 
brain failure it would seem to be as effective as WBRT when it comes to local
tumor control and overall survival, withdrawing the associated risks
within it \cite{Vlachos2023StereotacticMeta-analysis}.
\paragraph{}
Some early studies (Table 7) are indicative that surgery resection
combined with WBRT might reduce intracranial relapses and neurologic
deaths however it is not certain how it influences the duration of a
patient's functional independence since some studies present contrary
results regarding this matter. In terms of overall survival, resection
combined with WBRT does not reveal any advantages. A study by Hart et
al. \cite{Hart2005SurgicalMetastases} intended to assess the clinical 
effectiveness of surgical resection combined with WBRT versus WBRT 
alone in the treatment of patients with a single BM. The research 
comprehended three randomized controlled trials (RCTs) involving a 
total of 195 patients. Results showed that although surgery plus 
WBRT did not significantly improve overall survival compared to 
WBRT alone, it did however increase the duration of functionally 
independent survival (FIS). Nevertheless, this
study is contradicted by later research by Kocher et al. 
\cite{Kocher2011AdjuvantStudy} who
stated that though adjuvant WBRT decreases neurologic fatalities and
intracranial relapses, it does not extend functional independence or
increase overall survival. In this analysis the authors set out to
determine whether adjuvant WBRT could lengthen the period of functional
independence following BMs radiosurgery or resection. 359 patients
(n=199 underwent SRS; n=160 underwent resection) with 1 to 3 BMs
originated from solid tumors were randomly assigned to adjuvant WBRT or
observation. According to the data, adjuvant WBRT reduced intracranial
relapse rates (59\% to 27\% and 42\% to 23\%, at initial and new sites,
respectively) and neurologic deaths regardless, it did not improve the
duration of functional independence (approximately 10 months for both
WBRT and observation groups) or overall survival (roughly 11 months for
both WBRT and observation groups).
\paragraph{}
More recent studies have shown that the use of post-operative WBRT
resulted in a shorter OS and did not provide as much of local tumor
control as other methodologies of treatment such as SRS or FSRT.
Moreover, WBRT reveals to accelerate the decline of cognitive function
compared to other treatments. Research by Shafie et al. \cite{ElShafie2020StereotacticPatterns}
compared the use of single-session SRS or fractionated SRS (FSRS) of the
resection cavity versus WBRT in treating BMs after surgery. 101 patients
participated in the study, 50 of which were treated with SRS/FSRS and 51
were treated with WBRT. Findings indicated that patients who received
SRS/FSRT had longer OS and local control compared to those who received
WBRT. Following this analysis, Koo et al. \cite{Koo2021Whole-BrainStudy} retrospectively investigated which factors affected OS, local and distant control, and leptomeningeal metastases (LMM) in patients who had undergone resection of BMs. The study gathered 124 patients who had undergone resection of
BM, and results showed that although both WBRT and localized
radiotherapy (local brain radiotherapy or stereotactic radiosurgery
(LBRT/SRS)) demonstrated great local control, LBRT/SRS would still be
preferable to avoid WBRT effects on cognitive function and promote an
extended survival of cancer patients. Ultimately, research suggests that
after resection, alternative treatment options such as SRS may be better
for managing brain metastases compared to adjuvant WBRT.

\begin{longtable}{|p{2cm}|p{1cm}|p{4cm}|p{2cm}|p{5.4cm}|}
\captionsetup{singlelinecheck=false, justification=raggedright}
\caption{Studies and respective clinical outcomes regarding the application of Surgery Resection plus Whole Brain Radiation Therapy (WBRT) as treatment of BMs.} \\
    \hline
    \multicolumn{1}{|c|}{\textbf{Study}} &
    \multicolumn{1}{c|}{\textbf{Year}} &
    \multicolumn{1}{c|}{\textbf{Purpose}} &
    \multicolumn{1}{c|}{\textbf{Arm}} &
    \multicolumn{1}{c|}{\textbf{Clinical Outcomes}} \\
    \hline
    \endfirsthead
    \hline \multicolumn{5}{r}{{(Continues on next page)}} \\ 
    \endfoot
    \hline
    \endlastfoot
    Hart et al. \cite{Hart2005SurgicalMetastases} & 2005 & Assess the clinical effectiveness of surgical resection combined with WBRT versus WBRT alone in the treatment of patients with a single BM & Resection plus WBRT WBRT alone & Surgery plus WBRT did not significantly improve OS compared to WBRT alone but  increased the duration of functionally independent survival \\
    \hline
    Kocher et al. \cite{Kocher2011AdjuvantStudy} & 2011 &Determine whether adjuvant WBRT could lengthen the period of functional independence following 1 to 3 BMs originated from solid tumors & Resection plus WBRT & Adjuvant WBRT reduced intracranial relapse rates and neurologic deaths; resection plus WBRT did not improve the duration of functional independence or OS \\
    \hline
    Shafie et al. \cite{ElShafie2020StereotacticPatterns} & 2020 & Compare the use of single-session SRS or FSRT of the resection cavity versus WBRT in treating BMs after surgery & Resection plus single-session SRS Resection plus FSRS Resection plus WBRT & Patients who received SRS/FSRS had longer OS and local control compared to those who received WBRT \\
    \hline
    Koo et al. \cite{Koo2021Whole-BrainStudy} & 2021 & Investigate which factors affected OS, local and distant control, and leptomeningeal metastases (LMM) in patients who undergo resection of BMs & Resection plus SRS Resection plus WBRT &	After resection, SRS may be better for managing brain metastases compared to adjuvant WBRT \\
    \hline
\end{longtable}

\subsection{
\emph{Surgery Resection plus SRS}}
\paragraph{}
Several published data suggests that resection combined with SRS is an
effective treatment for patients with multiple metastases, tending
towards longer survival outcomes \cite{Mahajan2022StereotacticAnalysis}. Although 
WBRT might reduce the recurrence rate of BMs following resection, it does not 
increment overall survival neither quality of life, since WBRT is associated with
outcome toxicities namely, neurocognitive decline. This way, surgical
resection when combined with SRS is able to achieve great local control,
while preventing cognitive defects \cite{Patel2014ManagementMetastasis, 
Quigley2008SingleDisease, Mahajan2017Post-operativeTrial}.
An earlier study conducted by Mahajan et al. 
\cite{Mahajan2017Post-operativeTrial} suggests that SRS after BMs 
resection could potentially be more 
beneficial than WBRT, withdrawing the cognitive decline associated with it. The 
study intended to determine if SRS applied to the surgical cavity after resection 
could improve time to local recurrence compared with surgical resection alone.
Patients were selected by presenting complete resection of 1-3 BMs (maximum 
diameter of the resection cavity $\leq$4 cm), among other characteristics. The 
findings indicated that resection boosted with SRS significantly lowers local 
recurrence compared to resection alone, with a 12-month freedom from local 
recurrence of 72\% for the SRS group and 43\% for the observation group. 
Comparing to WBRT, surgical resection followed by SRS could not only prevent 
local recurrence, but also provide longer quality of life to patients lowering 
the hazard of cognitive decline. Additionally, the previously stated study by 
Shafie et al. \cite{ElShafie2020StereotacticPatterns} supports the fact that 
using SRS or fractioned SRS after surgery resection promotes longer OS and local 
control. The enhancement of local tumor control by irradiating the surgical 
cavity was also recently proven by Huang et al. \cite{Huang2023GammaMutations}. 
This retrospective study examined the effects of postoperative SRS on local tumor 
control and overall survival in 97 patients diagnosed with BMs who underwent Gamma
Knife SRS. The study found that irradiating the surgical cavity after
resection could enhance local tumor control rate by 75\% and overall
survival rate (12 months) by near 90\%. Hence, combining surgical
resection with SRS not only lowers local recurrence but also promotes
tumor volume control.
\paragraph{}
Despite the advancements in surgical care of brain tumors, the complete
resection of lesions located near vital neurovascular structures still
poses a significant risk of morbidity. Because it is minimally invasive,
SRS has low complication rates and excellent tumor control. However,
when it comes to larger tumor volumes, SRS alone does not constitute the
greatest treatment due to major radiation-induced problems. In recent
years, there has been growing interest in Adaptive Hybrid Surgery (AHS)
(BrainLab\textsuperscript{®}, Munich, Germany) as a multimodal method to
manage these large lesions. In AHS, a planned subtotal resection (STR)
with adjuvant SRS for an anticipated residual tumor takes advantage of
both strategies' benefits \cite{Cohen-Inbar2018AdaptiveNeurosurgery, 
Kienzler2022AdaptiveTumors, Sheppard2018PlannedSurgery}. Research by Sheppard et
al. \cite{Sheppard2018PlannedSurgery} compared the ideal radiosurgical target 
volumes defined by a manual method (surgeon) to those determined by Adaptive 
Hybrid Surgery (AHS) software in 7 patients with vestibular schwannoma (VS). The
findings showed that the planned residual tumor volumes were smaller
than the ideal radiosurgical target volumes defined by AHS (1.6
cm\textsuperscript{3} vs. 4.5 cm\textsuperscript{3}, respectively) and
the average difference between the ideal radiosurgical target volume
defined by AHS and the planned residual tumor volume was 2.9 ± 1.7
cm\textsuperscript{3}. Despite the fact that planned subtotal resection
of VS by a surgeon differs from the ideal radiosurgical target defined
by AHS, the study did not provide any favorable conclusion regarding the
preferred method since both methods influence in clinical outcomes were
not defined. Another survey by Cohen-Inbar et al. \cite{Cohen-Inbar2018AdaptiveNeurosurgery} favors
Adaptive Hybrid Surgery (AHS) as a combined strategy for the treatment
of large cerebral and skull base tumors. The study suggests that a
planned STR followed by SRS to a preplanned residual tumor offers a
desirable combination of high tumor control rates and favorable clinical
outcomes, including preservation of neurological function and quality of
life. Moreover, it recommends that patients with large tumors who cannot
be treated with SRS lone due to neurotoxicity outcomes, should be
managed with AHS. Recent research by Keinzler et al. \cite{Kienzler2022AdaptiveTumors} evaluated
the feasibility and safety of the Adaptive Hybrid Surgery Analysis
(AHSA) method in 5 patients with benign skull base tumors. Patients
underwent planned partial tumor resection followed by SRS. The AHSA
method was able to accurately assess the residual tumor volume during
surgery and suggest safe hypo-fractionated radiation plans in all
patients. No complications occurred after radiation treatment, and the
authors attested the viability of this technology and its great
potential in facilitating an optimal multidisciplinary approach and
resection strategy, reducing surgical and radiosurgical risks.
\paragraph{}
Ultimately, surgical resection combined with SRS might be recommended
for patients who have 1-3 brain metastases that are relatively small and
located in areas of the brain that can be safely accessed through
surgery (Table 8). The decision to use resection plus SRS will depend on
a number of factors, including the size and location of the tumors, the
patient\textquotesingle s overall health and prognosis, and the
potential risks and benefits of the treatment approach \cite{Patel2014ManagementMetastasis, Mahajan2017Post-operativeTrial, Huang2023GammaMutations}.

\begin{longtable}{|p{2cm}|p{1cm}|p{4cm}|p{2cm}|p{5.4cm}|}
\captionsetup{singlelinecheck=false, justification=raggedright}
\caption{Studies and respective clinical outcomes regarding the application of Surgery Resection plus Stereotactic Radiosurgery (SRS) as treatment of BMs.}\\
    \hline
    \multicolumn{1}{|c|}{\textbf{Study}} &
    \multicolumn{1}{c|}{\textbf{Year}} &
    \multicolumn{1}{c|}{\textbf{Purpose}} &
    \multicolumn{1}{c|}{\textbf{Arm}} &
    \multicolumn{1}{c|}{\textbf{Clinical Outcomes}} \\
    \hline
    \endfirsthead
    \hline
    \endhead
    \hline \multicolumn{5}{r}{{(Continues on next page)}} \\ 
    \endfoot
    \endlastfoot
    Mahajan et al. \cite{Mahajan2017Post-operativeTrial} & 2017 & Determine if 
    SRS applied to the surgical cavity of 1 to 3 BMs after resection could 
    improve time to local recurrence compared with surgical resection alone & 
    Resection alone Resection plus SRS & Resection boosted with SRS lowered local 
    recurrence compared to resection alone; 12-month freedom from local 
    recurrence was 72\% in the SRS group and 43\% in the observation group \\
    \hline
    Huang et al. \cite{Huang2023GammaMutations} & 2023 & Examine the effects of 
    postoperative SRS on local tumor control and overall survival in patients 
    diagnosed with BMs who underwent Gamma Knife SRS & Resection plus SRS & 
    Irradiating the surgical cavity after resection could enhance local tumor 
    control rate by 75\% and OS rate (12 months) by near 90\% \\
    \hline
\end{longtable}

\section{
\emph{Follow-up Care after Treatment of Multiple 
metastases}}
\paragraph{}
Even with successful treatment, BMs often recur, so medical
professionals recommend close follow-up after treatment. Individuals who
have BMs are at significant risk of experiencing recurrence of the tumor
in the same area or the development of new lesions in other parts of the
brain (distant recurrence). It is important to detect any potential
recurrence of the disease, even asymptomatic, to receive prompt
treatment. Follow-up care should be tailored to the individual patient
and based on factors such as the patient\textquotesingle s overall
health, the extent of metastases, and the type of treatment initially
received. Medical imaging in assistance to follow-up care is an
essential tool for evaluating the effectiveness of treatment and
distinguishing between any changes resulting from the treatment and
those indicating regrowth of the tumor \cite{UK2018Follow-upMetastases}.
\paragraph{}
Monitoring patients after treatment include regular imaging to detect
the progression of the disease and potential recurrence of new
metastases. Based on each patient's circumstances, the doctor will
determine the frequency imaging should take place. An experiment
executed by Jena et al. \cite{Jena2008MagneticSymptom} concluded that imaging 
surveillance of the brain could detect asymptomatic metastases. The study 
revealed that the pattern of the lesion may influence the patients' response to therapy and survival benefit, especially the asymptomatic patients with
multiple metastases. The research included 175 patients with BMs derived
from primary non-small cell lung cancer (NSCLC). Patients with
asymptomatic BMs and those with symptomatic BMs, were parted into two
separate groups. Results showed that more than 30\% of patients had BMs,
and roughly 47\% of those were asymptomatic. Also, no difference was
found in terms of number, size, site, nature, presence of perilesional
edema, and intralesional hemorrhage of the metastasis. Results showed
that the number of patients with and without neurological symptoms
caused by BMs, was roughly the same. Medical imaging during follow-up of
the brain, supports early detection of asymptomatic metastases though
the prognosis for these patients doesn't uniquely depend on the presence
or absence of symptoms.
\paragraph{}
A more recent survey by Derks et al. \cite{Derks2022BrainImaging} discussed the challenges posed by BMs and the role of imaging in clinical practice. The authors begin by outlining the rising prevalence of BMs from various primary
tumors and draw attention to the difficulty in choosing the correct
individuals for screening because not all cancer patients experience BM
development. The paper then goes into the imaging methods that are
employed to find BMs. A three-dimensional (3D) T1W MRI sequence is the
gold standard for BM detection, but other anatomical, functional, and
metabolic data from imaging methods like susceptibility weighted
imaging, diffusion weighted imaging, perfusion MRI, MR spectroscopy, and
positron emission tomography (PET) can help distinguish BMs from other
intracranial conditions. The authors also discuss the function of
imaging at various stages of BM treatment. For surgical resection,
imaging is utilized to choose surgical candidates and provide
intraoperative assistance using methods like fluorescence-guided surgery
and ultrasound. MRI and CT are combined for SRS therapy planning. After
both local and systemic therapy, conventional MRI is utilized for
surveillance, but sophisticated imaging is increasingly used to identify
real tumor progression.
\paragraph{}
Overall, effective follow-up care is an important part of the management
of patients with multiple BMs. Regular imaging, symptom management,
supportive care and patient education should be incorporated into an
individualized follow-up care plan. Follow-up care after treatment for
detecting new metastases, still represents a clinical challenge since
predicting the incidence of new metastases would be a great assistance
to medical professionals. This step forward in oncology would allow
radiologists and physicians to elaborate earlier secondary treatment
plans and study with more time alternatives to manage the newly
potential metastases.

\section{
\emph{Guidelines from multiple societies}}
\paragraph{}
Brain metastases are a serious complication of primary cancer, with an
estimated incidence of 10-30\% among cancer patients. Although there are
several treatment options available, the management of BMs is complex,
requiring a multidisciplinary approach and their optimal management
remains a topic of debate, with no clear consensus on the best approach
\cite{Amsbaugh2022BrainMetastasis, HowBrainlab.org, EpidemiologyUpToDate}. Multiple societies have developed guidelines for the
treatment and follow-up of multiple BMs, aiming to provide
evidence-based recommendations for clinicians on the best course of
action for managing patients with this condition, emphasizing the
importance of individualized treatment, considering factors such as the
patient's age, overall health status, and the extent of their cancer.
These guidelines typically cover a range of topics, including the most
effective treatments for BMs and how to monitor patients for recurrence
or new metastases. Because they are based on the available scientific
literature, these recommendations are updated periodically to reflect
new evidence or changes in clinical practice. Some of the societies that
have developed guidelines for the treatment and monitoring of multiple
BMs include the European Association of Neuro-Oncology (EANO) 
\cite{EANONeuro-Oncology}, the European Society for Medical Oncology (ESMO) 
\cite{EuropeanOncology}, the American Society for Radiation Oncology (ASTRO) 
\cite{HomeASTRO}, the American Society of
Clinical Oncology (ASCO) \cite{ASCOOncology} and the Korean Society for
Neuro-Oncology (KSNO) \cite{::::::}. The following sections (condensed in Figure 
3) will review the guidelines developed by each of these societies and
highlight their recommendations for the treatment and follow-up of
multiple BMs.

\begin{figure}[ht]
\centering
\includegraphics[width=12cm, height=10cm]{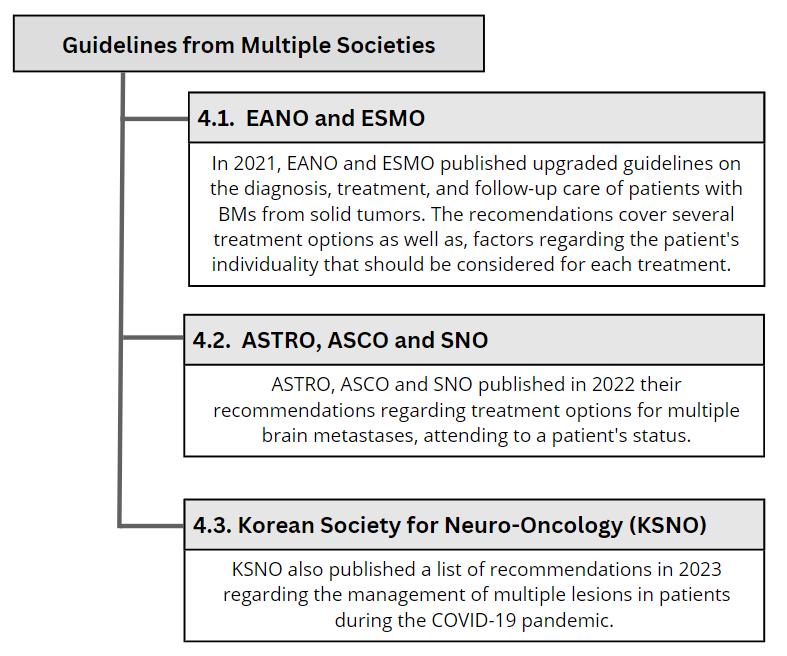}
\caption{Selection of the papers according to PRISMA diagram.}
\end{figure}

\subsection{
\emph{EANO and ESMO}}
\paragraph{}
The European Association of Neuro-Oncology (EANO) and the European
Society for Medical Oncology (ESMO) are two prominent professional
organizations that provide guidance on the management of BMs. EANO is a
multidisciplinary organization that brings together professionals from
various fields, to continuously launch new guidelines supporting the
development of high-quality care of brain tumor patients across Europe
\cite{GuidelinesCommittee}. The European Society for Medical Oncology (ESMO) is a leading organization for medical oncology professionals in Europe and around the world, with several missions including the improvement of cancer care from different sights, quality of prevention, diagnosis, treatment,
supportive and palliative care, as well as the follow-up of patients
with malignant diseases \cite{ESMOMission}. Based on an extensive examination of the scientific literature and professional opinion, EANO and ESMO have
both developed guidelines for the management of BMs.
\paragraph{}
In a 2017 publication \cite{Soffietti2017DiagnosisEANO} EANO presented evidence-based guidelines to help manage patients with BMs from solid tumors. These guidelines were created by a multidisciplinary Task Force and covered various aspects of diagnosis, staging, prognostic factors, and different
treatment options, including surgery, SRT, WBRT, chemotherapy, and
targeted therapy. The study provides recommendations for the management
of newly diagnosed BMs and specific attention to BMs from non-small cell
lung cancer, melanoma, breast cancer, and renal cancer, as well as
supportive care. A summary of these recommendations is presented on
Table 9.

\begin{table}[htbp]
    \captionsetup{singlelinecheck=false, justification=raggedright}
    \caption{EANO and ESMO guidelines for management of newly diagnosed brain metastases.}
    \begin{tabular}{|p{3.5cm}|p{\dimexpr\textwidth-3.5cm-4\tabcolsep-2\arrayrulewidth\relax}|}
        \hline
        \textbf{Treatment Modality} & \textbf{Recommendations} \\
        \hline
        \multirow{5}{*}{Surgical Resection} & Patients with 1 to 3 newly 
        diagnosed BMs, particularly lesions $\geq$3 cm in diameter, necrotic or 
        cystic appearance and edema/mass effect, lesions located in the 
        posterior fossa with associated hydrocephalus, and lesions located in 
        symptomatic eloquent areas \\
        \cline{2-2}
        & The systemic disease is absent/controlled and the KPS $\geq$60, as it can prolong survival \\
        \cline{2-2}
        & The systemic disease is active but there are effective systemic treatment options or when the primary tumor presents resistance to radiation \\
        \hline
        Stereotactic Fractionated Radiotherapy (SFRT) & Patients with metastases $>$3 cm in maximum diameter and a larger irradiation volume than 10 or 12 cm$^3$ due to increased toxicity and radiation necrosis of normal brain tissue \\
        \hline
        Stereotactic Radiosurgery (SRS) & Patients with metastases less than 3 to 3.5 cm \\
        \hline
        WBRT or best supportive care & Patients with short life expectancy (low KPS score and/or progressive systemic disease) \\
        \hline
    \end{tabular}
\end{table}
\paragraph{}
The recommendations also suggest that SRS and/or fractionated SRT should
be considered in patients with metastases that are not resectable due to
location or with other coexisting conditions. The use of adjuvant WBRT
after complete surgical resection or SRS is not unequivocally
recommended due to lack of survival advantage and risk of neurocognitive
decline and a close monitoring with MRI (every 3-4 months) is
recommended.
\paragraph{}
The decision regarding whether to employ surgical resection, SRS,
fractionated SRT, WBRT, alone or in combination, for patients with
multiple BMs comes down to clinical discretion, patient preference, and
logistical considerations. The study also refers to recommendations
regarding the treatment of recurrent BMs suggesting that resection must
be considered in selected patients with favorable prognostic factors and
accessible location, or when it is required a differential diagnosis
between tumor regrowth and radionecrosis. SRS followed by adjuvant WBRT
should be considered for local tumor control and OS and after an initial
course of SRS, additional rounds of SRS may be used as an alternative to
WBRT for new BMs.
\paragraph{}
In a more recent publication \cite{LeRhun2021EANO-ESMOTumours} both EANO and ESMO have stated upgraded recommendations on the diagnosis, treatment, and follow-up care of patients with BMs from solid tumors. Some of these guidelines recommend surgery resection when there is doubt on the neoplastic nature of a brain lesion (benign or malignant), when no primary tumor is known, when more than one tumour is known, or when there are acute symptoms of
raised intracranial pressure. Both single BMs and multiple resectable
BMs should be considered for resection and after surgery MRI should be
carried out within 48 hours to determine the extent of resection. SRS is
recommended for patients with 1 to 4 BMs and could be considered for
patients with 5 to 10 BMs with tumor volume \textless{} 15mL. SRS to the
resection cavity is recommended after complete or incomplete resection
of BMs. Post-operative WBRT after neurosurgical resection or after SRS
should be discouraged, and WBRT should only be considered to treat
multiple BMs that are not eligible for SRS, depending on the
symptomatology, size, number, and location of BMs, among others.
Moreover, supportive care with omission of WBRT should be considered in
patients with multiple BMs not eligible for SRS and poor preservation of
neurological status.
\paragraph{}
Regarding follow-up care the same study \cite{LeRhun2021EANO-ESMOTumours} 
suggests that every 2-3 months, when radiological progression is suspected or 
neurological symptoms develop, a detailed neurological examination should be done. Moreover, during monitoring the neurocognitive function and the
patient's ability to consent should be regularly assessed.

\subsection{
\emph{ASTRO, ASCO and SNO}}
\paragraph{}
The American Society for Radiation Oncology (ASTRO), the American
Society of Clinical Oncology (ASCO) and Society of Neuro Oncology (SNO)
are three prominent medical societies that focus on advancing the field
of oncology and improving cancer care for patients. ASTRO \cite{AboutASTRO} is a professional organization of physicians, nurses, biologists, physicists,
radiation therapists, dosimetrists and other health care professionals
who specialize in treating patients with radiation therapies. They are
dedicated in advancing the practice of radiation oncology by promoting
excellence in patient care, providing opportunities for educational and
professional development, promoting research, and disseminating research
results and representing radiation oncology in a rapidly evolving health
care environment. ASCO \cite{AboutASCO} is a professional organization of
physicians and oncology professionals who specialize in treating people
with cancer. ASCO is dedicated to advancing the field of oncology and
ensuring that everyone with cancer receives high-quality, equitable
care. SNO \cite{About} is a multidisciplinary organization of
neuro-oncologists, neurosurgeons, radiation oncologists, basic and
translational scientists, neuropathologists, nurses, and other health
care professionals, dedicated to promoting advances in neuro-oncology
through research and education. In a paper published in 2022 \cite{Vogelbaum2022TreatmentGuideline}
several guidelines for treatment of BMs were presented by these three
societies, ASCO, ASTRO and SNO. The recommendations regarding several
treatments are summarized in Table 10.

\begin{table}[htbp]
  \captionsetup{singlelinecheck=false, justification=raggedright}
  \caption{ASCO-ASTRO-SNO guidelines for management of newly diagnosed brain metastases.}
  \begin{tabular}{|>{\centering\arraybackslash}p{5cm}|p{\dimexpr\textwidth-5cm-4\tabcolsep-2\arrayrulewidth\relax}|}
    \hline
    \textbf{Treatment Modality} & \textbf{Recommendations} \\
    \hline
    \multirow{5}{*}{Surgical Resection} & Patients with suspected BMs without primary cancer diagnosis \\ 
    \cline{2-2}
    & Patients with large tumors with mass effect \\ 
    \cline{2-2}
    & Patients with multiple BMs and/or uncontrolled systemic disease are less likely to benefit from surgery unless the remaining disease is controllable via other measures \\
    \hline
    \multirow{5}{*}{Stereotactic radiosurgery (SRS)} & SRS alone (as opposed to WBRT or combination of WBRT and SRS) should be offered to patients with 1 to 4 unresected BMs, excluding small-cell carcinoma \\ 
    \cline{2-2}
    & Patients with 1 to 2 resected BMs if the surgical cavity can be safely treated and considering the extent of remaining intracranial disease \\
    \hline
  \end{tabular}
\end{table}
\paragraph{}
The guidelines also propose that when resection is considered as
treatment, no recommendation regarding the method of resection can be
made. Regarding symptomatic BMs, local therapy (SRS and/or radiation
therapy and/or surgery) should be the course of treatment and for
asymptomatic metastases, local therapy should not be postponed unless it
is specifically recommended. Concerning radiotherapy treatment, it should not be administered to patients with asymptomatic metastases who have KPS $\leq$50 or less, or KPS $<$70 and no systemic therapy options.
For patients with more than 4 unresected or more than 2 resected
metastases and improved performance status (KPS $>$70), SRS, WBRT, and the
combination of the two are all viable treatments.
\paragraph{}
More recently, ASTRO \cite{Gondi2022RadiationGuideline} published a new survey updating clinical practice guidelines for treating BMs with radiation therapy. The paper recommendations suggest SRS for patients with limited BMs and Eastern Cooperative Oncology Group performance status 0 to 2, and suggests
multidisciplinary discussion to consider surgical resections for all
tumors causing mass effect and/or that are \textgreater{} 4 cm. It also
recommends upfront local therapy for patients with symptomatic
metastases and SRS for those with resected BMs to improve local control.
The study also highlights the importance of multidisciplinary and
patient-centered decision-making for asymptomatic patients eligible for
central nervous system-active systemic therapy. Finally, the survey
strongly recommends early palliative care for patients with poor
prognosis. At the same time, ASCO released a paper \cite{Schiff2022RadiationGuideline} endorsing
the above work by ASTRO. The ASCO Endorsement Panel found the
recommendations to be clear and based on relevant scientific evidence.
The guidelines recommend the use of SRS for patients with up to 4 intact
BMs and with Eastern Cooperative Oncology Group performance status of
0-2. For patients with resected metastases, SRS, or WBRT is recommended
to improve intracranial tumor control, with SRS recommended over WBRT if
there are limited additional metastases. For patients with favorable
prognosis and BMs not eligible for surgery and/or SRS, WBRT with
hippocampal avoidance and memantine is recommended.

\subsection{
\emph{Korean Society for Neuro-Oncology (KSNO)}}
\paragraph{}
The Korean Society for Neuro-Oncology (KSNO) is a professional medical
society in South Korea, dedicated to the study and treatment of brain
and spinal cord tumors. This society gathers neurosurgeons,
neuro-oncologists, radiation oncologists, and other healthcare
professionals who specialize in the diagnosis and treatment of brain and
spinal cord tumors aiming to improve patient's outcomes through
education, research, and clinical practice guidelines. KSNO has recently
published a paper \cite{Kim2023The2023.1} describing treatment guidelines for patients with brain tumors during the COVID-19 pandemic. The society developed
the guidelines using the Delphi method with a working group of 22
multidisciplinary experts on neuro-oncology in Korea. The guidelines
prioritize surgery and radiotherapy based on appropriate time
window-based criteria for management outcome. The highest priority for
surgery is assigned to patients who are life-threatening or have a risk
of significant impact on prognosis unless immediate intervention is
given within 24 to 48 hours. Patients at risk of compromising their OS
or neurological status within 4 to 6 weeks are assigned to the highest
priority for radiotherapy. The study emphasizes the importance of
maintaining and providing adequate medical care for brain tumor patients
during times of crisis and provides a valuable resource for the delivery
of treatment to brain tumor patients in future crises.

\section{
\emph{Predicting Brain Metastases Incidence after RT treatment}}
\paragraph{}
Historically, being diagnosed with intracranial metastases meant that
life expectancy would be up to a month. With the introduction of WBRT,
however, life expectancy rose to 4-6 months \cite{Liu2019ManagementPresent}. 
Although there has been an exponential development in treatments of BMs, 
predicting their recurrence following a primary tumour still represents a 
challenge in oncology. Discriminating radiation necrosis from local recurrence 
(LR) or tumor progression remains one of the most challenging tasks.
Radiomics has demonstrated great promise in forecasting the likelihood
of recurrence or progression. These endpoints can be measured by
examination of radiomic features collected from pre-treatment imaging,
such as MRI or CT scans \cite{AbdelRazek2021ClinicalImaging, 
Ce2023ArtificialMedicine, Salvestrini2022TheyAIRO, 
Nowakowski2022RadiomicsMetastases}.
\subsection{
\emph{State-of-the-art technologies for predicting the incidence of brain 
metastases}}
\paragraph{}
Research on predicting the hazard of BMs after 
treatment is ongoing, and the state of the art is 
continually changing. Due to the novelty of this
topic, there haven't been many studies exploring 
this matter, however, research on the application 
of radiomics and machine learning algorithms
regarding the management of BMs is ongoing \cite{AbdelRazek2021ClinicalImaging, 
Ce2023ArtificialMedicine, Salvestrini2022TheyAIRO, 
Nowakowski2022RadiomicsMetastases, Zhao2018MachineSignature, Kim2022APatients}, 
and the approaches so far have significant potential for helping to treat metastases to the brain more successfully. The artificial intelligence discipline of radiomics enables the extraction of features from several modalities of medical images, followed by the creation of predictive models \cite{Salvestrini2022TheyAIRO}. 
When performing radiomic studies, the associations between pixels or voxels
in an image are quantified under the assumption that slight variations
in pixel/voxel intensity, location, and density can act as prognostic
and predictive biomarkers. The radiomics workflow follows five main
steps beginning with image acquisition, followed by image segmentation,
highlighting the region of interest, image preprocessing, feature
extraction, and ultimately clinical outcome prediction and
classification \cite{Nowakowski2022RadiomicsMetastases}.
\paragraph{}
A retrospective survey by Wang et al. \cite{Wang2018PredictingFeatures} 
investigated the potential of MRI-based radiomics features in predicting 
local recurrence of BMs after SRS. 18 patients who underwent SRS were 
included in the analysis and both pre-treatment and follow-up post-contrast 
T1-weighted MR images were used for radiomics analysis. Multiple radiomics 
features regarding image intensity and texture were extracted from the MRIs 
using PyRadiomics software. Logistic regression and receiver operating
characteristic (ROC) analysis were performed to evaluate the association
of these features with LR. The study also assessed the complementary
value of post-SRS features for LR prognosis. The findings showed that
tumor size was significantly associated with LR. Thirty pre-SRS
radiomics features demonstrated significant associations with LR.
According to the study\textquotesingle s findings, MRI radiomics
features before and after treatment can be used to predict the local
recurrence of BMs treated with SRS. Compared to using only pre-SRS
features, using post-SRS measures increased prediction accuracy. More
recently, retrospective research by Mulford et al. \cite{Mulford2021ASRS} 
assessed the predictive capability of radiomic-based models compared to 
standard clinical features for local tumor control in patients who received
adjuvant SRS for resected BMs. Radiomic features extracted from
pre-radiosurgery MRI scans of the surgical cavities were used to train a
gradient boosting model with K-fold cross validation. Results showed
that the radiomic-based model provided more robust predictions of local
control rates (AUC of 0.73) compared to models based solely on clinical
features (AUC of 0.40). In the end, the study found that compared to
clinical features alone, radiomics features offered more reliable
prediction models of local control rates. Similar to the earlier study
by Wang et al., the created model also showed improvements in
performance over clinical characteristics alone for predicting local
recurrence after adjuvant SRS to resected brain metastases. Radiomics
analysis of MR images can potentially offer useful information for
predicting LR in BMs after SRS, thus assisting in treatment planning and
patient care. The clinical value of MRI-based radiomics in this
situation needs to be established, nevertheless, and more study with
larger patient cohorts is required.
\paragraph{}
Traditional prognostic models frequently rely on pathological and
clinical variables. However, they might not adequately account for the
complexity of the lesions. The use of machine learning (ML) and deep
learning (DL) to improve the precision of recurrence
prediction in this context has gained popularity in recent years. 
An early survey by Zhao et al. \cite{Zhao2018MachineSignature} intended to find brain metastasis-related microRNAs (miRNAs) in lung adenocarcinoma (LUAD) and create a predictive model using a random Forest supervised classification algorithm. The researchers screened miRNA expression profiles from 77 LUAD patients with and without brain metastasis. Results showed that the predictive model presented high accuracy in stratifying patients into two groups with different brain metastasis sub types (roughly 90\%  and 91\% accuracy in the training and test sets, respectively). Moreover, the three miRNAs revealed to be an independent prediction factor closely connected with brain metastasis. 
\paragraph{}
In order to predict local failure in patients with BMs receiving hypo-fractionated SRT, Jaberipour et al. \cite{Jaberipour2021ALearning} tested the ability of pre-treatment quantitative MRI, clinical characteristics, and ML algorithms. Predictive models were created using data from 100 patients, and they were then tested on a separate test set utilizing data from 20 patients. The contrast-enhanced T1w and T2-FLAIR images used for treatment planning were used to extract quantitative MRI radiomic characteristics and clinical features were used to predict therapy outcome.
The developed quantitative MRI biomarker consisted of four features related to tumor heterogeneity and morphology. The predictive models with radiomic and clinical feature sets achieved AUCs of roughly 0.90 and 0.60, respectively, on the independent test set. Combining radiomic features with clinical features improved the model's performance by up to 16\%. The study's results showed a significant difference in survival between patient cohorts identified at pre-treatment using the radiomics-based predictive model and at post-treatment using standard clinical criteria. Findings demonstrate the potential of quantitative MRI radiomic features in predicting local failure in BMs treated with SRT, contributing to the advancement of precision oncology in this field.
\paragraph{}
Kim et al. \cite{Kim2022APatients} has recently published a survey developing 
a ML algorithm capable of predicting the possibility of BMs in patients with 
renal cell carcinoma (RCC), this way avoiding detection delays. The cohort 
for the research gathered 1282 RCC patients who were submitted to comparison 
in terms of the performance of six machine learning algorithms. The
adaptive boosting (AdaBoost) model outperformed the others with an area
under the receiver operating characteristic (AUROC) curve of 0.716. The
developed predictive model can predict the hazard of BMs incidence in
RCC patients and potentially help avoid detection delays by performing
CT scans on potentially asymptomatic patients.
\paragraph{}
Radiomics and ML models show great results in predicting the recurrence
of BMs following treatment. After SRS, the forecasting of BMs can be
achieved through MRI-based radiomics features, and accuracy is increased
by including post-SRS data. Although traditional approaches are
outperformed by ML models, this approach is yet to be further
investigated. There are currently few works regarding the application of
ML and DL models in forecasting the incidence of brain metastases after
primary treatment, during monitoring, which indicates its relatively new
application in this field of oncology.

\begin{longtable}{|m{2cm}|m{1cm}|m{4cm}|m{2cm}|m{5.4cm}|}
\captionsetup{singlelinecheck=false, justification=raggedright}
\caption{Technologies for predicting the incidence of brain metastases.}\\
    \hline
    \multicolumn{1}{|c|}{\textbf{Study}} &
    \multicolumn{1}{c|}{\textbf{Year}} &
    \multicolumn{1}{c|}{\textbf{Purpose}} &
    \multicolumn{1}{c|}{\textbf{Technology}} &
    \multicolumn{1}{c|}{\textbf{Clinical Outcomes}} \\
    \hline
    \endfirsthead
    \hline
    \endhead
    \hline \multicolumn{5}{r}{{(Continues on next page)}} \\ 
    \endfoot
    \endlastfoot
    Wang et al. \cite{Wang2018PredictingFeatures} & 2018 & Investigate the potential of MRI-based radiomics features in predicting local recurrence of BMs after SRS & PyRadiomics Software & Tumor size was significantly associated with LR; 30 pre-SRS radiomics features showed significant association with LR; MRI radiomics features before and after treatment could be used to predict the local recurrence of BMs treated with SRS \\
    \hline
    Mulford et al. \cite{Mulford2021ASRS} & 2021 & Compare the predictive capability of radiomic-based models to standard clinical features for LC in patients who received adjuvant SRS for resected BMs & Radiomics + gradient boosting model & The radiomic-based model provided more robust predictions of LC rates compared to models based solely on clinical features (AUC of 0.73 vs 0.40, respectively) \\
    \hline
    Zhao et al. \cite{Zhao2018MachineSignature} & 2018 & Analyze miRNA expression patterns to find BMs-related miRNAs and create a predictive model in tumor tissues using microarray technology & Random Forest supervised classification algorithm & The predictive model demonstrated great performance in stratifying patients into two groups with different brain metastasis sub types with accuracy of with 90\% and 91\% in training and test sets, respectively\\    
    \hline
    Jaberipour et al. \cite{Jaberipour2021ALearning} & 2021 & Predict local failure in patients with BMs receiving hypo-fractionated SRT & k-nearest neighbor (k-NN) & Predictive models with radiomic and clinical feature sets achieved AUCs of roughly 0.90 and 0.60, respectively. Combining radiomic features with clinical features improved the model's performance by up to 16\% \\   
    \hline
    Kim et al. \cite{Kim2022APatients} & 2022 & Develop a ML algorithm capable of predicting the possibility of BMs in patients with RCC & ML (AdaBoost) & The AdaBoost model outperformed the others with AUROC curve of 0.716 \\
    \hline
\end{longtable}

\subsection{
\emph{Implications for clinical practice}}
\paragraph{}
Predicting the risk of BMs after treatment through models and algorithms
brings numerous benefits, one of the most significant being the early
detection, which allows doctors to monitor patients closely and
intervene promptly to prevent or delay the development of BMs. Moreover,
early detection would help doctors to have different approaches towards
treatment planning and making the necessary alterations according to the
progression of the disease. The employment of predictive models in this
field of oncology, could also improve the patients' quality of life. The
developing of BMs can gradually impact a patient\textquotesingle s day
to day life, so the ability to predict the risk of developing these
metastases could assist patients and their families to prepare for the
upcoming changes and adjust their expectations accordingly. This can
reduce anxiety and stress related to the uncertainty of the
disease\textquotesingle s progression and help patients make informed
decisions about their care. Overall, the use of predictive models and
algorithms has several advantages, mainly improving patient's survival
outcomes and quality of life. Further research could lead to better
diagnosis, treatment, and patient care for those at risk of developing
BMs.

\section{
\emph{Conclusion}}
\paragraph{}
The incidence of brain metastases is increasingly denoted around the world due to ongoing improvement in treatment technology. The optimal management of newly diagnosed metastases is often controversial especially regarding the use of WBRT as primary or adjuvant treatment, due to the neurocognitive toxicities and poor overall survival related to it. Recommendations from several societies commonly relate to management of a limited number of BMs with surgical resection, employing SRS as adjuvant treatment to the surgical
cavity, improving control. Regarding monitorization of metastases,
follow-up care also remains a clinical challenge as the suitable period
for imaging is not universal. The employment of artificial intelligence models to assist medical professionals in the task of monitoring the incidence of new metastases or controlling the progression of the lesions, could anticipate several medical decisions and expand time of
analysis by employing predictive models attending post treatment images.
Despite conducting a comprehensive search across various databases, the literature review regarding this topic remains relatively scarce. The significance of brain metastases prediction in healthcare management cannot be understated. The paucity of research in this field may be attributed to the complex nature of the topic and the recent emergence of predictive technologies.

\printbibliography

@misc{::::::,
    title = {{::: THE KOREAN SOCIETY FOR RADIATION ONCOLOGY (KOSRO) :::}},
    url = {http://eng.kosro.or.kr/about/}
}

@article{Sahgal2020ACanada,
    title = {{A Cancer Care Ontario Organizational Guideline for the Delivery of Stereotactic Radiosurgery for Brain Metastasis in Ontario, Canada}},
    year = {2020},
    journal = {Practical radiation oncology},
    author = {Sahgal, Arjun and Kellett, Sarah and Ruschin, Mark and Greenspoon, Jeffrey and Follwell, Matthew and Sinclair, John and Islam, Omar and Perry, James},
    number = {4},
    month = {7},
    pages = {243--254},
    volume = {10},
    publisher = {Pract Radiat Oncol},
    url = {https://pubmed.ncbi.nlm.nih.gov/31783171/},
    doi = {10.1016/J.PRRO.2019.11.002},
    issn = {1879-8519},
    pmid = {31783171}
}

@article{Kim2022APatients,
    title = {{A Machine Learning Approach to Predict the Probability of Brain Metastasis in Renal Cell Carcinoma Patients}},
    year = {2022},
    journal = {Applied Sciences 2022, Vol. 12, Page 6174},
    author = {Kim, Hyung Min and Jeong, Chang Wook and Kwak, Cheol and Song, Cheryn and Kang, Minyong and Seo, Seong Il and Kim, Jung Kwon and Lee, Hakmin and Chung, Jinsoo and Hwang, Eu Chang and Park, Jae Young and Choi, In Young and Hong, Sung Hoo},
    number = {12},
    month = {6},
    pages = {6174},
    volume = {12},
    publisher = {Multidisciplinary Digital Publishing Institute},
    url = {https://www.mdpi.com/2076-3417/12/12/6174/htm https://www.mdpi.com/2076-3417/12/12/6174},
    doi = {10.3390/APP12126174},
    issn = {2076-3417}
}

@article{ElShafie2020AMetastases,
    title = {{A matched-pair analysis comparing stereotactic radiosurgery with whole-brain radiotherapy for patients with multiple brain metastases}},
    year = {2020},
    journal = {Journal of neuro-oncology},
    author = {El Shafie, Rami A. and Celik, Aylin and Weber, Dorothea and Schmitt, Daniela and Lang, Kristin and K{\"{o}}nig, Laila and Bernhardt, Denise and H{\"{o}}ne, Simon and Forster, Tobias and von Nettelbladt, Bastian and Adeberg, Sebastian and Debus, Jürgen and Rieken, Stefan},
    number = {3},
    month = {5},
    pages = {607--618},
    volume = {147},
    publisher = {J Neurooncol},
    url = {https://pubmed.ncbi.nlm.nih.gov/32239433/},
    doi = {10.1007/S11060-020-03447-2},
    issn = {1573-7373},
    pmid = {32239433}
}

@article{Jaberipour2021ALearning,
    title = {{A priori prediction of local failure in brain metastasis after hypo-fractionated stereotactic radiotherapy using quantitative MRI and machine learning}},
    year = {2021},
    journal = {Scientific Reports 2021 11:1},
    author = {Jaberipour, Majid and Soliman, Hany and Sahgal, Arjun and Sadeghi-Naini, Ali},
    number = {1},
    month = {11},
    pages = {1--10},
    volume = {11},
    publisher = {Nature Publishing Group},
    url = {https://www.nature.com/articles/s41598-021-01024-9},
    isbn = {0123456789},
    doi = {10.1038/s41598-021-01024-9},
    issn = {2045-2322},
    pmid = {34732781}
}

@article{Mulford2021ASRS,
    title = {{A radiomics-based model for predicting local control of resected brain metastases receiving adjuvant SRS}},
    year = {2021},
    journal = {Clinical and Translational Radiation Oncology},
    author = {Mulford, Kellen and Chen, Chuyu and Dusenbery, Kathryn and Yuan, Jianling and Hunt, Matthew A. and Chen, Clark C. and Sperduto, Paul and Watanabe, Yoichi and Wilke, Christopher},
    month = {7},
    pages = {27--32},
    volume = {29},
    publisher = {Elsevier},
    doi = {10.1016/J.CTRO.2021.05.001},
    issn = {2405-6308}
}

@misc{About,
    title = {{About}},
    url = {https://www.soc-neuro-onc.org/WEB/About/WEB/About.aspx?hkey=4a508251-f3f5-4336-8612-7c1fef6877a2}
}

@misc{AboutASCO,
    title = {{About ASCO | ASCO}},
    url = {https://old-prod.asco.org/about-asco}
}

@misc{AboutASTRO,
    title = {{About ASTRO - American Society for Radiation Oncology (ASTRO) - American Society for Radiation Oncology (ASTRO)}},
    url = {https://www.astro.org/About-ASTRO.aspx}
}

@article{Kienzler2022AdaptiveTumors,
    title = {{Adaptive Hybrid Surgery Experiences in Benign Skull Base Tumors}},
    year = {2022},
    journal = {Brain sciences},
    author = {Kienzler, Jenny Christine and Fandino, Javier},
    number = {10},
    month = {10},
    volume = {12},
    publisher = {Brain Sci},
    url = {https://pubmed.ncbi.nlm.nih.gov/36291260/},
    doi = {10.3390/BRAINSCI12101326},
    issn = {2076-3425},
    pmid = {36291260}
}

@article{Cohen-Inbar2018AdaptiveNeurosurgery,
    title = {{Adaptive Hybrid Surgery: Paradigm Shift for Patient-centered Neurosurgery}},
    year = {2018},
    journal = {Rambam Maimonides medical journal},
    author = {Cohen-Inbar, Or and Sviri, Gil E.},
    number = {3},
    month = {7},
    pages = {e0025},
    volume = {9},
    publisher = {Rambam Maimonides Med J},
    url = {https://pubmed.ncbi.nlm.nih.gov/30089092/},
    doi = {10.5041/RMMJ.10346},
    issn = {2076-9172},
    pmid = {30089092}
}

@article{Kocher2011AdjuvantStudy,
    title = {{Adjuvant Whole-Brain Radiotherapy Versus Observation After Radiosurgery or Surgical Resection of One to Three Cerebral Metastases: Results of the EORTC 22952-26001 Study}},
    year = {2011},
    journal = {Journal of Clinical Oncology},
    author = {Kocher, Martin and Soffietti, Riccardo and Abacioglu, Ufuk and Vill{\`{a}}, Salvador and Fauchon, Francois and Baumert, Brigitta G. and Fariselli, Laura and Tzuk-Shina, Tzahala and Kortmann, Rolf Dieter and Carrie, Christian and Ben Hassel, Mohamed and Kouri, Mauri and Valeinis, Egils and Van Den Berge, Dirk and Collette, Sandra and Collette, Laurence and Mueller, Rolf Peter},
    number = {2},
    month = {1},
    pages = {134},
    volume = {29},
    publisher = {American Society of Clinical Oncology},
    url = {/pmc/articles/PMC3058272/ /pmc/articles/PMC3058272/?report=abstract https://www.ncbi.nlm.nih.gov/pmc/articles/PMC3058272/},
    doi = {10.1200/JCO.2010.30.1655},
    issn = {0732183X},
    pmid = {21041710}
}

@article{Ding2023AnMetastases,
    title = {{An exploratory analysis of MR-guided fractionated stereotactic radiotherapy in patients with brain metastases}},
    year = {2023},
    journal = {Clinical and translational radiation oncology},
    author = {Ding, Shouliang and Liu, Biaoshui and Zheng, Shiyang and Wang, Daquan and Liu, Mingzhi and Liu, Hongdong and Zhang, Pengxin and Peng, Kangqiang and He, Haoqiang and Zhou, Rui and Guo, Jinyu and Qiu, Bo and Huang, Xiaoyan and Liu, Hui},
    month = {5},
    volume = {40},
    publisher = {Clin Transl Radiat Oncol},
    url = {https://pubmed.ncbi.nlm.nih.gov/36910023/},
    doi = {10.1016/J.CTRO.2023.100602},
    issn = {2405-6308},
    pmid = {36910023}
}

@article{Ce2023ArtificialMedicine,
    title = {{Artificial Intelligence in Brain Tumor Imaging: A Step toward Personalized Medicine}},
    year = {2023},
    journal = {Current oncology (Toronto, Ont.)},
    author = {C{\`{e}}, Maurizio and Irmici, Giovanni and Foschini, Chiara and Danesini, Giulia Maria and Falsitta, Lydia Viviana and Serio, Maria Lina and Fontana, Andrea and Martinenghi, Carlo and Oliva, Giancarlo and Cellina, Michaela},
    number = {3},
    month = {2},
    pages = {2673--2701},
    volume = {30},
    publisher = {Curr Oncol},
    url = {https://pubmed.ncbi.nlm.nih.gov/36975416/},
    doi = {10.3390/CURRONCOL30030203},
    issn = {1718-7729},
    pmid = {36975416}
}

@misc{ASCOOncology,
    title = {{ASCO Hub – American Society of Clinical Oncology}},
    url = {https://www.asco.org/}
}

@article{Liu2022BrainIncidence,
    title = {{Brain Metastases among Cancer Patients Diagnosed from 2010-2017 in Canada: Incidence Proportion at Diagnosis and Estimated Lifetime Incidence}},
    year = {2022},
    journal = {Current oncology (Toronto, Ont.)},
    author = {Liu, Jiaqi L. and Walker, Emily V. and Paudel, Yuba Raj and Davis, Faith G. and Yuan, Yan},
    number = {3},
    month = {3},
    pages = {2091--2105},
    volume = {29},
    publisher = {Curr Oncol},
    url = {https://pubmed.ncbi.nlm.nih.gov/35323369/},
    doi = {10.3390/CURRONCOL29030169},
    issn = {1718-7729},
    pmid = {35323369}
}

@article{Fabi2011BrainCenter,
    title = {{Brain metastases from solid tumors: Disease outcome according to type of treatment and therapeutic resources of the treating center}},
    year = {2011},
    journal = {Journal of Experimental and Clinical Cancer Research},
    author = {Fabi, Alessandra and Felici, Alessandra and Metro, Giulio and Mirri, Alessandra and Bria, Emilio and Telera, Stefano and Moscetti, Luca and Russillo, Michelangelo and Lanzetta, Gaetano and Mansueto, Giovanni and Pace, Andrea and Maschio, Marta and Vidiri, Antonello and Sperduti, Isabella and Cognetti, Francesco and Carapella, Carmine M.},
    number = {1},
    month = {1},
    pages = {1--7},
    volume = {30},
    publisher = {BioMed Central},
    url = {https://jeccr.biomedcentral.com/articles/10.1186/1756-9966-30-10},
    doi = {10.1186/1756-9966-30-10/TABLES/5},
    issn = {17569966},
    pmid = {21244695}
}

@article{Popat2021BrainEra,
    title = {{Brain metastases in solid tumours: new guidelines for a new era}},
    year = {2021},
    journal = {Annals of Oncology},
    author = {Popat, S. and Welsh, L.},
    number = {11},
    month = {11},
    pages = {1322--1324},
    volume = {32},
    publisher = {Elsevier Ltd},
    url = {http://www.annalsofoncology.org/article/S0923753421042770/fulltext http://www.annalsofoncology.org/article/S0923753421042770/abstract https://www.annalsofoncology.org/article/S0923-7534(21)04277-0/abstract},
    doi = {10.1016/j.annonc.2021.08.1992},
    issn = {15698041},
    pmid = {34455070}
}

@article{Nieder2023BrainTherapy,
    title = {{Brain Metastases: Is There Still a Role for Whole-Brain Radiation Therapy?}},
    year = {2023},
    journal = {Seminars in Radiation Oncology},
    author = {Nieder, Carsten and Andratschke, Nicolaus H. and Grosu, Anca L.},
    month = {4},
    publisher = {W.B. Saunders},
    doi = {10.1016/J.SEMRADONC.2023.01.005},
    issn = {15329461}
}

@misc{BrainReview,
    title = {{Brain metastases: Literature review}},
    url = {https://www.elsevier.es/en-revista-revista-medica-del-hospital-general-325-pdf-S0185106316300166}
}

@article{Derks2022BrainImaging,
    title = {{Brain metastases: the role of clinical imaging}},
    year = {2022},
    journal = {The British journal of radiology},
    author = {Derks, Sophie H.A.E. and van der Veldt, Astrid A.M. and Smits, Marion},
    number = {1130},
    volume = {95},
    publisher = {Br J Radiol},
    url = {https://pubmed.ncbi.nlm.nih.gov/34808072/},
    doi = {10.1259/BJR.20210944},
    issn = {1748-880X},
    pmid = {34808072}
}

@article{Amsbaugh2022BrainMetastasis,
    title = {{Brain Metastasis}},
    year = {2022},
    journal = {StatPearls},
    author = {Amsbaugh, Mark J. and Kim, Catherine S.},
    month = {4},
    publisher = {StatPearls Publishing},
    url = {https://www.ncbi.nlm.nih.gov/books/NBK470246/},
    pmid = {29262201}
}

@article{AbdelRazek2021ClinicalImaging,
    title = {{Clinical applications of artificial intelligence and radiomics in neuro-oncology imaging}},
    year = {2021},
    journal = {Insights into Imaging 2021 12:1},
    author = {Abdel Razek, Ahmed Abdel Khalek and Alksas, Ahmed and Shehata, Mohamed and AbdelKhalek, Amr and Abdel Baky, Khaled and El-Baz, Ayman and Helmy, Eman},
    number = {1},
    month = {10},
    pages = {1--17},
    volume = {12},
    publisher = {SpringerOpen},
    url = {https://insightsimaging.springeropen.com/articles/10.1186/s13244-021-01102-6},
    doi = {10.1186/S13244-021-01102-6},
    issn = {1869-4101}
}

@article{Ogawa2022ClinicalSurgery,
    title = {{Clinical Outcomes and Prognostic Factors of Fractionated Stereotactic Radiosurgery for Brain Metastases >=20 mm as a Potential Alternative to Surgery}},
    year = {2022},
    journal = {World neurosurgery},
    author = {Ogawa, Hiroaki and Ito, Kei and Karasawa, Katsuyuki},
    month = {6},
    pages = {e141-e146},
    volume = {162},
    publisher = {World Neurosurg},
    url = {https://pubmed.ncbi.nlm.nih.gov/35247616/},
    doi = {10.1016/J.WNEU.2022.02.098},
    issn = {1878-8769},
    pmid = {35247616}
}

@article{Mizuno2019ComparisonCancer,
    title = {{Comparison between stereotactic radiosurgery and whole-brain radiotherapy for 10-20 brain metastases from non-small cell lung cancer}},
    year = {2019},
    journal = {Molecular and clinical oncology},
    author = {Mizuno, Takaaki and Takada, Kazuto and Hasegawa, Toshinori and Yoshida, Tatsuya and Murotani, Kenta and Kobayashi, Hironori and Sakurai, Tsutomu and Yamashita, Yuuki and Akazawa, Nana and Kojima, Eiji},
    number = {5},
    month = {3},
    volume = {10},
    publisher = {Mol Clin Oncol},
    url = {https://pubmed.ncbi.nlm.nih.gov/30967951/},
    doi = {10.3892/MCO.2019.1830},
    issn = {2049-9450},
    pmid = {30967951}
}

@article{Khan2017ComparisonMeta-analysis,
    title = {{Comparison of WBRT alone, SRS alone, and their combination in the treatment of one or more brain metastases: Review and meta-analysis}},
    year = {2017},
    author = {Khan, Muhammad and Lin, Jie and Liao, Guixiang and Li, Rong and Wang, Baiyao and Xie, Guozhu and Zheng, Jieling and Yuan, Yawei},
    url = {https://doi.org/10.1177/1010428317702903},
    doi = {10.1177/1010428317702903}
}

@article{Sittenfeld2018ContemporaryMetastases,
    title = {{Contemporary management of 1-4 brain metastases}},
    year = {2018},
    journal = {Frontiers in Oncology},
    author = {Sittenfeld, Sarah M.C. and Suh, John H. and Murphy, Erin S. and Yu, Jennifer S. and Chao, Samuel T.},
    number = {SEP},
    month = {9},
    pages = {385},
    volume = {8},
    publisher = {Frontiers Media S.A.},
    doi = {10.3389/FONC.2018.00385/BIBTEX},
    issn = {2234943X}
}

@article{Suh2020CurrentMetastases,
    title = {{Current approaches to the management of brain metastases}},
    year = {2020},
    journal = {Nature Reviews Clinical Oncology 2020 17:5},
    author = {Suh, John H. and Kotecha, Rupesh and Chao, Samuel T. and Ahluwalia, Manmeet S. and Sahgal, Arjun and Chang, Eric L.},
    number = {5},
    month = {2},
    pages = {279--299},
    volume = {17},
    publisher = {Nature Publishing Group},
    url = {https://www.nature.com/articles/s41571-019-0320-3},
    doi = {10.1038/s41571-019-0320-3},
    issn = {1759-4782},
    pmid = {32080373}
}

@article{Saha2013DemographicStudy,
    title = {{Demographic and clinical profile of patients with brain metastases: A retrospective study}},
    year = {2013},
    journal = {Asian Journal of Neurosurgery},
    author = {Saha, Animesh and Ghosh, Sajal Kumar and Roy, Chhaya and Choudhury, Krishnangshu Bhanja and Chakrabarty, Bikramjit and Sarkar, Ratan},
    number = {3},
    month = {9},
    pages = {157},
    volume = {8},
    publisher = {Thieme Medical Publishers},
    url = {/pmc/articles/PMC3877503/ /pmc/articles/PMC3877503/?report=abstract https://www.ncbi.nlm.nih.gov/pmc/articles/PMC3877503/},
    doi = {10.4103/1793-5482.121688},
    issn = {1793-5482},
    pmid = {24403959}
}

@article{Soffietti2017DiagnosisEANO,
    title = {{Diagnosis and treatment of brain metastases from solid tumors: guidelines from the European Association of Neuro-Oncology (EANO)}},
    year = {2017},
    journal = {Neuro-oncology},
    author = {Soffietti, Riccardo and Abacioglu, Ufuk and Baumert, Brigitta and Combs, Stephanie E. and Kinhult, Sara and Kros, Johan M. and Marosi, Christine and Metellus, Philippe and Radbruch, Alexander and Freixa, Salvador S.Villa and Brada, Michael and Carapella, Carmine M. and Preusser, Matthias and Le Rhun, Emilie and Rud{\`{a}}, Roberta and Tonn, Joerg C. and Weber, Damien C. and Weller, Michael},
    number = {2},
    month = {2},
    pages = {162--174},
    volume = {19},
    publisher = {Neuro Oncol},
    url = {https://pubmed.ncbi.nlm.nih.gov/28391295/},
    doi = {10.1093/NEUONC/NOW241},
    issn = {1523-5866},
    pmid = {28391295}
}

@article{Chea2021DosimetricCharacteristics,
    title = {{Dosimetric study between a single isocenter dynamic conformal arc therapy technique and Gamma Knife radiosurgery for multiple brain metastases treatment: impact of target volume geometrical characteristics}},
    year = {2021},
    journal = {Radiation Oncology (London, England)},
    author = {Chea, Michel and Fezzani, Karen and Jacob, Julian and Cuttat, Marguerite and Crois{\'{e}}, Mathilde and Simon, Jean Marc and Feuvret, Loïc and Valery, Charles Ambroise and Maingon, Philippe and Benadjaoud, Mohamed Amine and Jenny, Catherine},
    number = {1},
    month = {12},
    pages = {45},
    volume = {16},
    publisher = {BioMed Central},
    url = {/pmc/articles/PMC7912819/ /pmc/articles/PMC7912819/?report=abstract https://www.ncbi.nlm.nih.gov/pmc/articles/PMC7912819/},
    doi = {10.1186/S13014-021-01766-W},
    issn = {1748717X},
    pmid = {33639959}
}

@misc{EANONeuro-Oncology,
    title = {{EANO - European Association of Neuro-Oncology}},
    url = {https://www.eano.eu/#}
}

@article{LeRhun2021EANO-ESMOTumours,
    title = {{EANO-ESMO Clinical Practice Guidelines for diagnosis, treatment and follow-up of patients with brain metastasis from solid tumours}},
    year = {2021},
    journal = {Annals of oncology : official journal of the European Society for Medical Oncology},
    author = {Le Rhun, E. and Guckenberger, M. and Smits, M. and Dummer, R. and Bachelot, T. and Sahm, F. and Galldiks, N. and de Azambuja, E. and Berghoff, A. S. and Metellus, P. and Peters, S. and Hong, Y. K. and Winkler, F. and Schadendorf, D. and van den Bent, M. and Seoane, J. and Stahel, R. and Minniti, G. and Wesseling, P. and Weller, M. and Preusser, M.},
    number = {11},
    month = {11},
    pages = {1332--1347},
    volume = {32},
    publisher = {Ann Oncol},
    url = {https://pubmed.ncbi.nlm.nih.gov/34364998/},
    doi = {10.1016/J.ANNONC.2021.07.016},
    issn = {1569-8041},
    pmid = {34364998}
}

@article{Brown2016EffectTrial,
    title = {{Effect of Radiosurgery Alone vs Radiosurgery With Whole Brain Radiation Therapy on Cognitive Function in Patients With 1 to 3 Brain Metastases: A Randomized Clinical Trial}},
    year = {2016},
    journal = {JAMA},
    author = {Brown, Paul D. and Jaeckle, Kurt and Ballman, Karla V. and Farace, Elana and Cerhan, Jane H. and Keith Anderson, S. and Carrero, Xiomara W. and Barker, Fred G. and Deming, Richard and Burri, Stuart H. and M{\'{e}}nard, Cynthia and Chung, Caroline and Stieber, Volker W. and Pollock, Bruce E. and Galanis, Evanthia and Buckner, Jan C. and Asher, Anthony L.},
    number = {4},
    month = {7},
    pages = {401},
    volume = {316},
    publisher = {NIH Public Access},
    url = {/pmc/articles/PMC5313044/ /pmc/articles/PMC5313044/?report=abstract https://www.ncbi.nlm.nih.gov/pmc/articles/PMC5313044/},
    doi = {10.1001/JAMA.2016.9839},
    issn = {15383598},
    pmid = {27458945}
}

@article{Stafinski2006EffectivenessMeta-analysis,
    title = {{Effectiveness of stereotactic radiosurgery alone or in combination with whole brain radiotherapy compared to conventional surgery and/or whole brain radiotherapy for the treatment of one or more brain metastases: A systematic review and meta-analysis}},
    year = {2006},
    journal = {Cancer Treatment Reviews},
    author = {Stafinski, Tania and Jhangri, Gian S. and Yan, Elizabeth and Menon, Devidas},
    number = {3},
    month = {5},
    pages = {203--213},
    volume = {32},
    publisher = {W.B. Saunders},
    doi = {10.1016/J.CTRV.2005.12.009},
    issn = {0305-7372},
    pmid = {16472924}
}

@article{Fox2011EpidemiologyTumors,
    title = {{Epidemiology of metastatic brain tumors}},
    year = {2011},
    journal = {Neurosurgery clinics of North America},
    author = {Fox, Benjamin D. and Cheung, Vincent J. and Patel, Akash J. and Suki, Dima and Rao, Ganesh},
    number = {1},
    month = {1},
    pages = {1--6},
    volume = {22},
    publisher = {Neurosurg Clin N Am},
    url = {https://pubmed.ncbi.nlm.nih.gov/21109143/},
    doi = {10.1016/J.NEC.2010.08.007},
    issn = {1558-1349},
    pmid = {21109143}
}

@article{Singh2020EpidemiologyMetastases,
    title = {{Epidemiology of synchronous brain metastases}},
    year = {2020},
    journal = {Neuro-oncology Advances},
    author = {Singh, Raj and Stoltzfus, Kelsey C. and Chen, Hanbo and Louie, Alexander V. and Lehrer, Eric J. and Horn, Samantha R. and Palmer, Joshua D. and Trifiletti, Daniel M. and Brown, Paul D. and Zaorsky, Nicholas G.},
    number = {1},
    month = {1},
    pages = {1--10},
    volume = {2},
    publisher = {Oxford University Press},
    url = {/pmc/articles/PMC7182307/ /pmc/articles/PMC7182307/?report=abstract https://www.ncbi.nlm.nih.gov/pmc/articles/PMC7182307/},
    doi = {10.1093/NOAJNL/VDAA041},
    issn = {26322498},
    pmid = {32363344}
}

@misc{EpidemiologyUpToDate,
    title = {{Epidemiology, clinical manifestations, and diagnosis of brain metastases - UpToDate}},
    url = {https://www.uptodate.com/contents/epidemiology-clinical-manifestations-and-diagnosis-of-brain-metastases}
}

@misc{ESMOMission,
    title = {{ESMO Mission}},
    url = {https://www.esmo.org/about-esmo/esmo-mission}
}

@misc{EuropeanOncology,
    title = {{European Society for Medical Oncology}},
    url = {https://www.esmo.org/}
}

@article{Cui2022EvaluationIsocenter,
    title = {{Evaluation of two automated treatment planning techniques for multiple brain metastases using a single isocenter}},
    year = {2022},
    journal = {Journal of Radiosurgery and SBRT},
    author = {Cui, Guoqiang and Yang, Yun and Yin, Fang-Fang and Yoo, David and Kim, Grace and Duan, Jun},
    number = {1},
    month = {11},
    pages = {47},
    volume = {8},
    publisher = {Old City Publishing},
    url = {/pmc/articles/PMC8930061/ /pmc/articles/PMC8930061/?report=abstract https://www.ncbi.nlm.nih.gov/pmc/articles/PMC8930061/},
    doi = {10.1016/j.ijrobp.2020.07.682},
    issn = {03603016},
    pmid = {35387403}
}

@article{Layer2023Five-FractionAnalysis,
    title = {{Five-Fraction Stereotactic Radiotherapy for Brain Metastases-A Retrospective Analysis}},
    year = {2023},
    journal = {Current oncology (Toronto, Ont.)},
    author = {Layer, Julian P. and Layer, Katharina and Sarria, Gustavo R. and R{\"{o}}hner, Fred and Dejonckheere, Cas S. and Friker, Lea L. and Zeyen, Thomas and Koch, David and Scafa, Davide and Leitzen, Christina and K{\"{o}}ksal, Mümtaz and Schmeel, Frederic Carsten and Sch{\"{a}}fer, Niklas and Landsberg, Jennifer and H{\"{o}}lzel, Michael and Herrlinger, Ulrich and Schneider, Matthias and Giordano, Frank A. and Schmeel, Leonard Christopher},
    number = {2},
    month = {2},
    pages = {1300--1313},
    volume = {30},
    publisher = {Curr Oncol},
    url = {https://pubmed.ncbi.nlm.nih.gov/36826062/},
    doi = {10.3390/CURRONCOL30020101},
    issn = {1718-7729},
    pmid = {36826062}
}

@article{Piras2022Five-FractionSchedules,
    title = {{Five-Fraction Stereotactic Radiotherapy for Brain Metastases: A Single-Institution Experience on Different Dose Schedules}},
    year = {2022},
    journal = {Oncology research and treatment},
    author = {Piras, Antonio and Boldrini, Luca and Menna, Sebastiano and Sanfratello, Antonella and D'Aviero, Andrea and Cusumano, Davide and Di Cristina, Luciana and Messina, Marco and Spada, Massimiliano and Angileri, Tommaso and Daidone, Antonino},
    number = {7-8},
    month = {8},
    pages = {408--414},
    volume = {45},
    publisher = {Oncol Res Treat},
    url = {https://pubmed.ncbi.nlm.nih.gov/35172322/},
    doi = {10.1159/000522645},
    issn = {2296-5262},
    pmid = {35172322}
}

@article{UK2018Follow-upMetastases,
    title = {{Follow-up for brain metastases}},
    year = {2018},
    author = {(UK), National Guideline Alliance},
    publisher = {National Institute for Health and Care Excellence (NICE)},
    url = {https://www.ncbi.nlm.nih.gov/books/NBK570047/}
}

@article{Baliga2017FractionatedModelling,
    title = {{Fractionated stereotactic radiation therapy for brain metastases: a systematic review with tumour control probability modelling}},
    year = {2017},
    journal = {The British journal of radiology},
    author = {Baliga, Sujith and Garg, Madhur K. and Fox, Jana and Kalnicki, Shalom and Lasala, Patrick A. and Welch, Mary R. and Tom{\'{e}}, Wolfgang A. and Ohri, Nitin},
    number = {1070},
    volume = {90},
    publisher = {Br J Radiol},
    url = {https://pubmed.ncbi.nlm.nih.gov/27936894/},
    doi = {10.1259/BJR.20160666},
    issn = {1748-880X},
    pmid = {27936894}
}

@article{Putz2020FSRTStudy,
    title = {{FSRT vs. SRS in Brain Metastases—Differences in Local Control and Radiation Necrosis—A Volumetric Study}},
    year = {2020},
    journal = {Frontiers in Oncology},
    author = {Putz, Florian and Weissmann, Thomas and Oft, Dominik and Schmidt, Manuel Alexander and Roesch, Johannes and Siavooshhaghighi, Hadi and Filimonova, Irina and Schmitter, Charlotte and Mengling, Veit and Bert, Christoph and Frey, Benjamin and Lettmaier, Sebastian and Distel, Luitpold Valentin and Semrau, Sabine and Fietkau, Rainer},
    month = {9},
    pages = {559193},
    volume = {10},
    publisher = {Frontiers Media S.A.},
    doi = {10.3389/FONC.2020.559193/BIBTEX},
    issn = {2234943X}
}

@article{Huang2023GammaMutations,
    title = {{Gamma Knife Radiosurgery Irradiation of Surgical Cavity of Brain Metastases: Factor Analysis and Gene Mutations}},
    year = {2023},
    journal = {Life (Basel, Switzerland)},
    author = {Huang, Yi Han and Yang, Huai Che and Chiang, Chi Lu and Wu, Hsiu Mei and Luo, Yung Hung and Hu, Yong Sin and Lin, Chung Jung and Chung, Wen Yuh and Shiau, Cheng Ying and Guo, Wan Yuo and Lee, Cheng Chia},
    number = {1},
    month = {1},
    volume = {13},
    publisher = {Life (Basel)},
    url = {https://pubmed.ncbi.nlm.nih.gov/36676186/},
    doi = {10.3390/LIFE13010236},
    issn = {2075-1729},
    pmid = {36676186}
}

@misc{GuidelinesCommittee,
    title = {{Guidelines Committee}},
    url = {https://www.eano.eu/about/about-us/committees/guidelines-committee/}
}

@misc{HomeASTRO,
    title = {{Home - American Society for Radiation Oncology (ASTRO)}},
    url = {https://www.astro.org/}
}

@misc{HowBrainlab.org,
    title = {{How Common Are Brain Metastases? - Brainlab.org}},
    url = {https://www.brainlab.org/get-educated/brain-metastasis/investigate-brain-metastasis/how-common-are-brain-metastases/}
}

@article{Habbous2020IncidenceStudy,
    title = {{Incidence and real-world burden of brain metastases from solid tumors and hematologic malignancies in Ontario: a population-based study}},
    year = {2020},
    journal = {Neuro-oncology advances},
    author = {Habbous, Steven and Forster, Katharina and Darling, Gail and Jerzak, Katarzyna and Holloway, Claire M.B. and Sahgal, Arjun and Das, Sunit},
    number = {1},
    month = {1},
    volume = {3},
    publisher = {Neurooncol Adv},
    url = {https://pubmed.ncbi.nlm.nih.gov/33585818/},
    doi = {10.1093/NOAJNL/VDAA178},
    issn = {2632-2498},
    pmid = {33585818}
}

@article{Salzmann2022Long-termImmunotherapy,
    title = {{Long-term neurocognitive function after whole-brain radiotherapy in patients with melanoma brain metastases in the era of immunotherapy}},
    year = {2022},
    journal = {Strahlentherapie und Onkologie},
    author = {Salzmann, Martin and Hess, Klaus and Lang, Kristin and Enk, Alexander H. and Jordan, Berit and Hassel, Jessica C.},
    number = {10},
    month = {10},
    pages = {884--891},
    volume = {198},
    publisher = {Springer Science and Business Media Deutschland GmbH},
    url = {https://link.springer.com/article/10.1007/s00066-022-01950-1},
    doi = {10.1007/S00066-022-01950-1/TABLES/2},
    issn = {1439099X},
    pmid = {35546362}
}

@article{Zhao2018MachineSignature,
    title = {{Machine Learning Based Prediction of Brain Metastasis of Patients with IIIA-N2 Lung Adenocarcinoma by a Three-miRNA Signature}},
    year = {2018},
    journal = {Translational Oncology},
    author = {Zhao, Shuangtao and Yu, Jiangyong and Wang, Luhua},
    pages = {157--167},
    volume = {11},
    url = {https://doi.org/10.1016/j.tranon.2017.12.002},
    doi = {10.1016/j.tranon.2017.12.002}
}

@article{Jena2008MagneticSymptom,
    title = {{Magnetic resonance (MR) patterns of brain metastasis in lung cancer patients: Correlation of imaging findings with symptom}},
    year = {2008},
    journal = {Journal of Thoracic Oncology},
    author = {Jena, Amarnath and Taneja, Sangeeta and Talwar, Vineet and Sharma, Jai Bhagwan},
    number = {2},
    month = {2},
    pages = {140--144},
    volume = {3},
    publisher = {International Association for the Study of Lung Cancer},
    url = {http://www.jto.org/article/S1556086415314118/fulltext http://www.jto.org/article/S1556086415314118/abstract https://www.jto.org/article/S1556-0864(15)31411-8/abstract},
    doi = {10.1097/JTO.0b013e318161d775},
    issn = {15561380},
    pmid = {18303434}
}

@article{Liu2019ManagementPresent,
    title = {{Management of brain metastases: history and the present}},
    year = {2019},
    journal = {Chinese Neurosurgical Journal},
    author = {Liu, Qi and Tong, Xuezhi and Wang, Jiangfei},
    number = {1},
    month = {1},
    volume = {5},
    publisher = {BioMed Central},
    url = {/pmc/articles/PMC7398203/ /pmc/articles/PMC7398203/?report=abstract https://www.ncbi.nlm.nih.gov/pmc/articles/PMC7398203/},
    doi = {10.1186/S41016-018-0149-0},
    issn = {20574967},
    pmid = {32922901}
}

@article{Patel2014ManagementMetastasis,
    title = {{Management of Cerebral Brain Metastasis}},
    year = {2014},
    journal = {Current Surgery Reports 2014 2:12},
    author = {Patel, Mira A. and Ruzevick, Jacob and Lim, Michael},
    number = {12},
    month = {10},
    pages = {1--7},
    volume = {2},
    publisher = {Springer},
    url = {https://link.springer.com/article/10.1007/s40137-014-0074-x},
    doi = {10.1007/S40137-014-0074-X},
    issn = {2167-4817}
}

@article{Lauko2020MedicalMetastases,
    title = {{Medical management of brain metastases}},
    year = {2020},
    journal = {Neuro-oncology Advances},
    author = {Lauko, Adam and Rauf, Yasmeen and Ahluwalia, Manmeet S.},
    number = {1},
    month = {1},
    volume = {2},
    publisher = {Oxford University Press},
    url = {/pmc/articles/PMC7415253/ /pmc/articles/PMC7415253/?report=abstract https://www.ncbi.nlm.nih.gov/pmc/articles/PMC7415253/},
    doi = {10.1093/NOAJNL/VDAA015},
    issn = {26322498},
    pmid = {32793881}
}

@article{Chang2009NeurocognitionTrial,
    title = {{Neurocognition in patients with brain metastases treated with radiosurgery or radiosurgery plus whole-brain irradiation: a randomised controlled trial}},
    year = {2009},
    journal = {The Lancet Oncology},
    author = {Chang, Eric L. and Wefel, Jeffrey S. and Hess, Kenneth R. and Allen, Pamela K. and Lang, Frederick F. and Kornguth, David G. and Arbuckle, Rebecca B. and Swint, J. Michael and Shiu, Almon S. and Maor, Moshe H. and Meyers, Christina A.},
    number = {11},
    month = {11},
    pages = {1037--1044},
    volume = {10},
    publisher = {Elsevier},
    doi = {10.1016/S1470-2045(09)70263-3},
    issn = {1470-2045},
    pmid = {19801201}
}

@article{Sheppard2018PlannedSurgery,
    title = {{Planned Subtotal Resection of Vestibular Schwannoma Differs from the Ideal Radiosurgical Target Defined by Adaptive Hybrid Surgery}},
    year = {2018},
    journal = {World neurosurgery},
    author = {Sheppard, John P. and Lagman, Carlito and Prashant, Giyarpuram N. and Alkhalid, Yasmine and Nguyen, Thien and Duong, Courtney and Udawatta, Methma and Gaonkar, Bilwaj and Tenn, Stephen E. and Bloch, Orin and Yang, Isaac},
    month = {6},
    pages = {e441-e446},
    volume = {114},
    publisher = {World Neurosurg},
    url = {https://pubmed.ncbi.nlm.nih.gov/29530701/},
    doi = {10.1016/J.WNEU.2018.03.005},
    issn = {1878-8769},
    pmid = {29530701}
}

@article{Mahajan2017Post-operativeTrial,
    title = {{Post-operative stereotactic radiosurgery versus observation for completely resected brain metastases: a single-centre, randomised, controlled, phase 3 trial}},
    year = {2017},
    journal = {The Lancet Oncology},
    author = {Mahajan, Anita and Ahmed, Salmaan and McAleer, Mary Frances and Weinberg, Jeffrey S. and Li, Jing and Brown, Paul and Settle, Stephen and Prabhu, Sujit S. and Lang, Frederick F. and Levine, Nicholas and McGovern, Susan and Sulman, Erik and McCutcheon, Ian E. and Azeem, Syed and Cahill, Daniel and Tatsui, Claudio and Heimberger, Amy B. and Ferguson, Sherise and Ghia, Amol and Demonte, Franco and Raza, Shaan and Guha-Thakurta, Nandita and Yang, James and Sawaya, Raymond and Hess, Kenneth R. and Rao, Ganesh},
    number = {8},
    month = {8},
    pages = {1040--1048},
    volume = {18},
    publisher = {Elsevier},
    doi = {10.1016/S1470-2045(17)30414-X},
    issn = {1470-2045},
    pmid = {28687375}
}

@article{Wang2018PredictingFeatures,
    title = {{Predicting Local Recurrence of Stereotactic Radiosurgery Brain Metastases Using MRI Radiomics Features}},
    year = {2018},
    journal = {International Journal of Radiation Oncology*Biology*Physics},
    author = {Wang, H. and Barbee, D. and Xue, J. and Das, I.J. and Kondziolka, D.},
    number = {3},
    month = {11},
    pages = {e563-e564},
    volume = {102},
    publisher = {Elsevier BV},
    url = {http://www.redjournal.org/article/S0360301618330189/fulltext http://www.redjournal.org/article/S0360301618330189/abstract https://www.redjournal.org/article/S0360-3016(18)33018-9/abstract},
    doi = {10.1016/j.ijrobp.2018.07.1564},
    issn = {03603016}
}

@misc{PRISMA,
    title = {{PRISMA}},
    url = {http://www.prisma-statement.org/PRISMAStatement/}
}

@article{Garsa2021RadiationReview,
    title = {{Radiation Therapy for Brain Metastases: A Systematic Review}},
    year = {2021},
    journal = {Practical radiation oncology},
    author = {Garsa, Adam and Jang, Julie K. and Baxi, Sangita and Chen, Christine and Akinniranye, Olamigoke and Hall, Owen and Larkin, Jody and Motala, Aneesa and Hempel, Susanne},
    number = {5},
    month = {9},
    pages = {354--365},
    volume = {11},
    publisher = {Pract Radiat Oncol},
    url = {https://pubmed.ncbi.nlm.nih.gov/34119447/},
    doi = {10.1016/J.PRRO.2021.04.002},
    issn = {1879-8519},
    pmid = {34119447}
}

@article{Gondi2022RadiationGuideline,
    title = {{Radiation Therapy for Brain Metastases: An ASTRO Clinical Practice Guideline}},
    year = {2022},
    journal = {Practical radiation oncology},
    author = {Gondi, Vinai and Bauman, Glenn and Bradfield, Lisa and Burri, Stuart H. and Cabrera, Alvin R. and Cunningham, Danielle A. and Eaton, Bree R. and Hattangadi‐Gluth, Jona A. and Kim, Michelle M. and Kotecha, Rupesh and Kraemer, Lianne and Li, Jing and Nagpal, Seema and Rusthoven, Chad G. and Suh, John H. and Tom{\'{e}}, Wolfgang A. and Wang, Tony J.C. and Zimmer, Alexandra S. and Ziu, Mateo and Brown, Paul D.},
    number = {4},
    month = {7},
    pages = {265--282},
    volume = {12},
    publisher = {Pract Radiat Oncol},
    url = {https://pubmed.ncbi.nlm.nih.gov/35534352/},
    doi = {10.1016/J.PRRO.2022.02.003},
    issn = {1879-8519},
    pmid = {35534352}
}

@article{Schiff2022RadiationGuideline,
    title = {{Radiation Therapy for Brain Metastases: ASCO Guideline Endorsement of ASTRO Guideline}},
    year = {2022},
    journal = {Journal of clinical oncology : official journal of the American Society of Clinical Oncology},
    author = {Schiff, David and Messersmith, Hans and Brastianos, Priscilla K. and Brown, Paul D. and Burri, Stuart and Dunn, Ian F. and Gaspar, Laurie E. and Gondi, Vinai and Jordan, Justin T. and Maues, Julia and Mohile, Nimish and Redjal, Navid and Stevens, Glen H.J. and Sulman, Erik P. and Van Den Bent, Martin and Wallace, H. James and Zadeh, Gelareh and Vogelbaum, Michael A.},
    number = {20},
    month = {5},
    volume = {40},
    publisher = {J Clin Oncol},
    url = {https://pubmed.ncbi.nlm.nih.gov/35561283/},
    doi = {10.1200/JCO.22.00333},
    issn = {1527-7755},
    pmid = {35561283}
}

@article{Nowakowski2022RadiomicsMetastases,
    title = {{Radiomics as an emerging tool in the management of brain metastases}},
    year = {2022},
    journal = {Neuro-Oncology Advances},
    author = {Nowakowski, Alexander and Lahijanian, Zubin and Panet-Raymond, Valerie and Siegel, Peter M. and Petrecca, Kevin and Maleki, Farhad and Dankner, Matthew},
    number = {1},
    month = {1},
    pages = {1--16},
    volume = {4},
    publisher = {Oxford Academic},
    url = {https://dx.doi.org/10.1093/noajnl/vdac141},
    doi = {10.1093/NOAJNL/VDAC141},
    issn = {26322498}
}

@article{Tsao2012RadiotherapeuticGuideline,
    title = {{Radiotherapeutic and surgical management for newly diagnosed brain metastasis(es): An American Society for Radiation Oncology evidence-based guideline}},
    year = {2012},
    journal = {Practical Radiation Oncology},
    author = {Tsao, May N. and Rades, Dirk and Wirth, Andrew and Lo, Simon S. and Danielson, Brita L. and Gaspar, Laurie E. and Sperduto, Paul W. and Vogelbaum, Michael A. and Radawski, Jeffrey D. and Wang, Jian Z. and Gillin, Michael T. and Mohideen, Najeeb and Hahn, Carol A. and Chang, Eric L.},
    number = {3},
    month = {7},
    pages = {210--225},
    volume = {2},
    publisher = {Elsevier},
    url = {http://www.practicalradonc.org/article/S1879850011003808/fulltext http://www.practicalradonc.org/article/S1879850011003808/abstract https://www.practicalradonc.org/article/S1879-8500(11)00380-8/abstract},
    doi = {10.1016/j.prro.2011.12.004},
    issn = {18798500}
}

@article{Che2022RecentExperience,
    title = {{Recent Trends in Synchronous Brain Metastasis Incidence and Mortality in the United States: Ten-Year Multicenter Experience}},
    year = {2022},
    journal = {Current oncology (Toronto, Ont.)},
    author = {Che, Wenqiang and Liu, Jie and Fu, Tengyue and Wang, Xiangyu and Lyu, Jun},
    number = {11},
    month = {11},
    pages = {8374--8389},
    volume = {29},
    publisher = {Curr Oncol},
    url = {https://pubmed.ncbi.nlm.nih.gov/36354720/},
    doi = {10.3390/CURRONCOL29110660},
    issn = {1718-7729},
    pmid = {36354720}
}

@article{Gong2022RecurrenceAdenocarcinoma,
    title = {{Recurrence benefit from supramarginal resection in brain metastases of lung adenocarcinoma}},
    year = {2022},
    journal = {Heliyon},
    author = {Gong, Weizhao and Jiang, Taipeng and Zuo, Dahui},
    number = {8},
    month = {8},
    pages = {e10109},
    volume = {8},
    publisher = {Elsevier},
    doi = {10.1016/J.HELIYON.2022.E10109},
    issn = {2405-8440}
}

@article{Heler2022RecurrentSetting,
    title = {{Recurrent brain metastases: the role of resection of in a comprehensive multidisciplinary treatment setting}},
    year = {2022},
    journal = {BMC Cancer},
    author = {He{\ss}ler, Nadine and J{\"{u}}nger, Stephanie T. and Meissner, Anna Katharina and Kocher, Martin and Goldbrunner, Roland and Grau, Stefan},
    number = {1},
    month = {12},
    pages = {275},
    volume = {22},
    publisher = {BioMed Central},
    url = {/pmc/articles/PMC8922794/ /pmc/articles/PMC8922794/?report=abstract https://www.ncbi.nlm.nih.gov/pmc/articles/PMC8922794/},
    doi = {10.1186/S12885-022-09317-6},
    issn = {14712407},
    pmid = {35291972}
}

@article{Nakano2022Reduced-doseControl,
    title = {{Reduced-dose WBRT combined with SRS for 1-4 brain metastases aiming at minimizing neurocognitive function deterioration without compromising brain tumor control}},
    year = {2022},
    journal = {Clinical and translational radiation oncology},
    author = {Nakano, Toshimichi and Aoyama, Hidefumi and Onodera, Shunsuke and Igaki, Hiroshi and Matsumoto, Yasuo and Kanemoto, Ayae and Shimamoto, Shigetoshi and Matsuo, Masayuki and Tanaka, Hidekazu and Oya, Natsuo and Matsuyama, Tomohiko and Ohta, Atsushi and Maruyama, Katsuya and Tanaka, Takahiro and Kitamura, Nobutaka and Akazawa, Kohei and Maebayashi, Katsuya},
    month = {11},
    pages = {116--129},
    volume = {37},
    publisher = {Clin Transl Radiat Oncol},
    url = {https://pubmed.ncbi.nlm.nih.gov/36199814/},
    doi = {10.1016/J.CTRO.2022.09.005},
    issn = {2405-6308},
    pmid = {36199814}
}

@article{Junger2021ResectionMetastases,
    title = {{Resection of symptomatic non-small cell lung cancer brain metastasis in the setting of multiple brain metastases}},
    year = {2021},
    journal = {Journal of neurosurgery},
    author = {J{\"{u}}nger, Stephanie T. and Reinecke, David and Meissner, Anna Katharina and Goldbrunner, Roland and Grau, Stefan},
    number = {6},
    month = {6},
    pages = {1576--1582},
    volume = {136},
    publisher = {J Neurosurg},
    url = {https://pubmed.ncbi.nlm.nih.gov/34715653/},
    doi = {10.3171/2021.7.JNS211172},
    issn = {1933-0693},
    pmid = {34715653}
}

@article{ChaoROENTGEN-RAYMETASTASES,
    title = {{ROENTGEN-RAY THERAPY OF CEREBRAL METASTASES}},
    author = {Chao, Jen-Hung and Phillips, Ralph and Nickson, James J},
    url = {https://onlinelibrary.wiley.com/terms-and-conditions},
    doi = {10.1002/1097-0142}
}

@article{Limon2017SingleMetastases,
    title = {{Single fraction stereotactic radiosurgery for multiple brain metastases}},
    year = {2017},
    journal = {Advances in Radiation Oncology},
    author = {Limon, Dror and McSherry, Frances and Herndon, James and Sampson, John and Fecci, Peter and Adamson, Justus and Wang, Zhiheng and Yin, Fang Fang and Floyd, Scott and Kirkpatrick, John and Kim, Grace J.},
    number = {4},
    month = {10},
    pages = {555--563},
    volume = {2},
    publisher = {Elsevier},
    doi = {10.1016/J.ADRO.2017.09.002},
    issn = {2452-1094}
}

@article{Quigley2008SingleDisease,
    title = {{Single session stereotactic radiosurgery boost to the post-operative site in lieu of whole brain radiation in metastatic brain disease}},
    year = {2008},
    journal = {Journal of neuro-oncology},
    author = {Quigley, Matthew R. and Fuhrer, Russell and Karlovits, Stephen and Karlovits, Brian and Johnson, Mark},
    number = {3},
    pages = {327--332},
    volume = {87},
    publisher = {J Neurooncol},
    url = {https://pubmed.ncbi.nlm.nih.gov/18183353/},
    doi = {10.1007/S11060-007-9515-Z},
    issn = {0167-594X},
    pmid = {18183353}
}

@article{Raza2022Single-isocenterSystems,
    title = {{Single-isocenter multiple-target stereotactic radiosurgery for multiple brain metastases: dosimetric evaluation of two automated treatment planning systems}},
    year = {2022},
    journal = {Radiation oncology (London, England)},
    author = {Raza, Giorgio Hamid and Capone, Luca and Tini, Paolo and Giraffa, Martina and Gentile, Piercarlo and Minniti, Giuseppe},
    number = {1},
    month = {12},
    volume = {17},
    publisher = {Radiat Oncol},
    url = {https://pubmed.ncbi.nlm.nih.gov/35778741/},
    doi = {10.1186/S13014-022-02086-3},
    issn = {1748-717X},
    pmid = {35778741}
}

@article{ElShafie2020StereotacticPatterns,
    title = {{Stereotactic Cavity Irradiation or Whole-Brain Radiotherapy Following Brain Metastases Resection—Outcome, Prognostic Factors, and Recurrence Patterns}},
    year = {2020},
    journal = {Frontiers in Oncology},
    author = {El Shafie, Rami A. and Dresel, Thorsten and Weber, Dorothea and Schmitt, Daniela and Lang, Kristin and K{\"{o}}nig, Laila and H{\"{o}}ne, Simon and Forster, Tobias and von Nettelbladt, Bastian and Eichkorn, Tanja and Adeberg, Sebastian and Debus, Jürgen and Rieken, Stefan and Bernhardt, Denise},
    month = {5},
    pages = {693},
    volume = {10},
    publisher = {Frontiers Media SA},
    url = {/pmc/articles/PMC7232539/ /pmc/articles/PMC7232539/?report=abstract https://www.ncbi.nlm.nih.gov/pmc/articles/PMC7232539/},
    doi = {10.3389/FONC.2020.00693},
    issn = {2234943X},
    pmid = {32477942}
}

@article{Harris2022StereotacticRadiosurgery,
    title = {{Stereotactic Radiosurgery}},
    year = {2022},
    journal = {StatPearls},
    author = {Harris, Lauren and Das, Joe M},
    month = {7},
    publisher = {StatPearls Publishing},
    url = {https://www.ncbi.nlm.nih.gov/books/NBK542166/ http://www.pubmedcentral.nih.gov/articlerender.fcgi?artid=PMC2556333},
    pmid = {31194323}
}

@article{Qie2018StereotacticTrials,
    title = {{Stereotactic radiosurgery (SRS) alone versus whole brain radiotherapy plus SRS in patients with 1 to 4 brain metastases from non-small cell lung cancer stratified by the graded prognostic assessment: A meta-analysis (PRISMA) of randomized control trials}},
    year = {2018},
    journal = {Medicine},
    author = {Qie, Shuai and Li, Yanhong and Shi, Hong Yun and Yuan, Lanhui and Su, Lei and Zhang, Xi},
    number = {33},
    month = {8},
    volume = {97},
    publisher = {Medicine (Baltimore)},
    url = {https://pubmed.ncbi.nlm.nih.gov/30113464/},
    doi = {10.1097/MD.0000000000011777},
    issn = {1536-5964},
    pmid = {30113464}
}

@article{Mahajan2022StereotacticAnalysis,
    title = {{Stereotactic radiosurgery and resection for treatment of multiple brain metastases: a systematic review and analysis}},
    year = {2022},
    journal = {Neurosurgical focus},
    author = {Mahajan, Uma V. and Desai, Ansh and Shost, Michael D. and Cai, Yang and Anthony, Austin and Labak, Collin M. and Herring, Eric Z. and Wijesekera, Olindi and Mukherjee, Debraj and Sloan, Andrew E. and Hodges, Tiffany R.},
    number = {5},
    volume = {53},
    publisher = {Neurosurg Focus},
    url = {https://pubmed.ncbi.nlm.nih.gov/36321293/},
    doi = {10.3171/2022.8.FOCUS22369},
    issn = {1092-0684},
    pmid = {36321293}
}

@article{Yamamoto2014StereotacticStudy,
    title = {{Stereotactic radiosurgery for patients with multiple brain metastases (JLGK0901): a multi-institutional prospective observational study}},
    year = {2014},
    journal = {The Lancet Oncology},
    author = {Yamamoto, Masaaki and Serizawa, Toru and Shuto, Takashi and Akabane, Atsuya and Higuchi, Yoshinori and Kawagishi, Jun and Yamanaka, Kazuhiro and Sato, Yasunori and Jokura, Hidefumi and Yomo, Shoji and Nagano, Osamu and Kenai, Hiroyuki and Moriki, Akihito and Suzuki, Satoshi and Kida, Yoshihisa and Iwai, Yoshiyasu and Hayashi, Motohiro and Onishi, Hiroaki and Gondo, Masazumi and Sato, Mitsuya and Akimitsu, Tomohide and Kubo, Kenji and Kikuchi, Yasuhiro and Shibasaki, Toru and Goto, Tomoaki and Takanashi, Masami and Mori, Yoshimasa and Takakura, Kintomo and Saeki, Naokatsu and Kunieda, Etsuo and Aoyama, Hidefumi and Momoshima, Suketaka and Tsuchiya, Kazuhiro},
    number = {4},
    month = {4},
    pages = {387--395},
    volume = {15},
    publisher = {Elsevier},
    doi = {10.1016/S1470-2045(14)70061-0},
    issn = {1470-2045},
    pmid = {24621620}
}

@article{Badiyan2016StereotacticMetastases,
    title = {{Stereotactic radiosurgery for treatment of brain metastases}},
    year = {2016},
    journal = {Journal of Oncology Practice},
    author = {Badiyan, Shahed N. and Regine, William F. and Mehta, Minesh},
    number = {8},
    month = {8},
    pages = {703--712},
    volume = {12},
    publisher = {American Society of Clinical Oncology},
    doi = {10.1200/JOP.2016.012922},
    issn = {1935469X},
    pmid = {27511715}
}

@article{Gaebe2022StereotacticMeta-analysis,
    title = {{Stereotactic radiosurgery versus whole brain radiotherapy in patients with intracranial metastatic disease and small-cell lung cancer: a systematic review and meta-analysis}},
    year = {2022},
    journal = {The Lancet. Oncology},
    author = {Gaebe, Karolina and Li, Alyssa Y. and Park, Amy and Parmar, Ambica and Lok, Benjamin H. and Sahgal, Arjun and Chan, Kelvin K.W. and Erickson, Anders W. and Das, Sunit},
    number = {7},
    month = {7},
    pages = {931--939},
    volume = {23},
    publisher = {Lancet Oncol},
    url = {https://pubmed.ncbi.nlm.nih.gov/35644163/},
    doi = {10.1016/S1470-2045(22)00271-6},
    issn = {1474-5488},
    pmid = {35644163}
}

@article{Vlachos2023StereotacticMeta-analysis,
    title = {{Stereotactic radiosurgery versus whole-brain radiotherapy after resection of solitary brain metastasis: A systematic review and meta-analysis}},
    year = {2023},
    journal = {World Neurosurgery: X},
    author = {Vlachos, Nikolaos and Lampros, Marios G. and Filis, Panagiotis and Voulgaris, Spyridon and Alexiou, George A.},
    month = {4},
    pages = {100170},
    volume = {18},
    publisher = {Elsevier},
    doi = {10.1016/J.WNSX.2023.100170},
    issn = {2590-1397}
}

@article{Chen2005Stereotactic1,
    title = {{Stereotactic Radiosurgery: Instrumentation and Theoretical Aspects—Part 1}},
    year = {2005},
    journal = {The Permanente Journal},
    author = {Chen, Joseph C T and Girvigian, Michael R},
    number = {4},
    month = {9},
    pages = {23},
    volume = {9},
    publisher = {Kaiser Permanente},
    url = {/pmc/articles/PMC3396107/ /pmc/articles/PMC3396107/?report=abstract https://www.ncbi.nlm.nih.gov/pmc/articles/PMC3396107/},
    doi = {10.7812/TPP/04-075},
    issn = {15525767},
    pmid = {22811641}
}

@article{Lupattelli2022StereotacticOligometastases,
    title = {{Stereotactic radiotherapy for brain oligometastases}},
    year = {2022},
    journal = {Reports of Practical Oncology and Radiotherapy},
    author = {Lupattelli, Marco and Tini, Paolo and Nardone, Valerio and Aristei, Cynthia and Borghesi, Simona and Maranzano, Ernesto and Anselmo, Paola and Ingrosso, Gianluca and Deantonio, Letizia and di Monale e Bastia, Michela Buglione},
    number = {1},
    pages = {15},
    volume = {27},
    publisher = {Via Medica sp. z o.o. sp. k.},
    url = {/pmc/articles/PMC8989457/ /pmc/articles/PMC8989457/?report=abstract https://www.ncbi.nlm.nih.gov/pmc/articles/PMC8989457/},
    doi = {10.5603/RPOR.A2021.0133},
    issn = {15071367},
    pmid = {35402029}
}

@article{Winther2022SurgeryResection,
    title = {{Surgery for brain metastases-impact of the extent of resection}},
    year = {2022},
    journal = {Acta neurochirurgica},
    author = {Winther, Rebecca Rootwelt and Hjermstad, Marianne Jensen and Skovlund, Eva and Aass, Nina and Helseth, Eirik and Kaasa, Stein and Yri, Olav Erich and Vik-Mo, Einar Osland},
    number = {10},
    month = {10},
    pages = {2773--2780},
    volume = {164},
    publisher = {Acta Neurochir (Wien)},
    url = {https://pubmed.ncbi.nlm.nih.gov/35080651/},
    doi = {10.1007/S00701-021-05104-7},
    issn = {0942-0940},
    pmid = {35080651}
}

@article{Ene2022SurgicalNuances,
    title = {{Surgical Management of Brain Metastasis: Challenges and Nuances}},
    year = {2022},
    journal = {Frontiers in oncology},
    author = {Ene, Chibawanye I. and Ferguson, Sherise D.},
    month = {3},
    volume = {12},
    publisher = {Front Oncol},
    url = {https://pubmed.ncbi.nlm.nih.gov/35359380/},
    doi = {10.3389/FONC.2022.847110},
    issn = {2234-943X},
    pmid = {35359380}
}

@article{Hart2005SurgicalMetastases,
    title = {{Surgical resection and whole brain radiation therapy versus whole brain radiation therapy alone for single brain metastases}},
    year = {2005},
    journal = {The Cochrane Database of Systematic Reviews},
    author = {Hart, Michael G. and Walker, Mark and Dickinson, Heather O. and Grant, Robin},
    number = {1},
    month = {1},
    volume = {2005},
    publisher = {John Wiley and Sons, Inc. and the Cochrane Library},
    url = {/pmc/articles/PMC6457740/ /pmc/articles/PMC6457740/?report=abstract https://www.ncbi.nlm.nih.gov/pmc/articles/PMC6457740/},
    doi = {10.1002/14651858.CD003292.PUB2},
    issn = {14651858},
    pmid = {15674905}
}

@article{Schodel2020SurgicalTreatment,
    title = {{Surgical resection of symptomatic brain metastases improves the clinical status and facilitates further treatment}},
    year = {2020},
    journal = {Cancer medicine},
    author = {Sch{\"{o}}del, Petra and J{\"{u}}nger, Stephanie T. and Wittersheim, Maike and Reinhardt, Hans Christian and Schmidt, Nils Ole and Goldbrunner, Roland and Proescholdt, Martin and Grau, Stefan},
    number = {20},
    month = {10},
    pages = {7503--7510},
    volume = {9},
    publisher = {Cancer Med},
    url = {https://pubmed.ncbi.nlm.nih.gov/32858763/},
    doi = {10.1002/CAM4.3402},
    issn = {2045-7634},
    pmid = {32858763}
}

@article{AmaanAliSurvivalLesions,
    title = {{Survival patterns of 5750 SRS-treated brain metastasis patients as a function of the number of lesions}},
    author = {Amaan Ali, Mir and Hirshman, Brian R and Wilson, Bayard and Carroll, Kate T and Proudfoot, James A and Goetsch, Steven J and Alksne, John F and Ott, Kenneth and Aiyama, Hitoshi and Nagano, Osamu and Carter, Bob S and Fogarty, Gerald and Hong, Angela and Serizawa, Toru and Yamamoto, Masaaki and Chen, Clark C},
    doi = {10.1016/j.wneu.2017.07.062}
}

@article{Mehta2005TheMetastases,
    title = {{The American Society for Therapeutic Radiology and Oncology (ASTRO) evidence-based review of the role of radiosurgery for brain metastases}},
    year = {2005},
    journal = {International journal of radiation oncology, biology, physics},
    author = {Mehta, Minesh P. and Tsao, May N. and Whelan, Timothy J. and Morris, David E. and Hayman, James A. and Flickinger, John C. and Mills, Michael and Rogers, C. Leland and Souhami, Luis},
    number = {1},
    month = {9},
    pages = {37--46},
    volume = {63},
    publisher = {Int J Radiat Oncol Biol Phys},
    url = {https://pubmed.ncbi.nlm.nih.gov/16111570/},
    doi = {10.1016/J.IJROBP.2005.05.023},
    issn = {0360-3016},
    pmid = {16111570}
}

@article{Yao2023TheRadiotherapy,
    title = {{The correlations between psychological distress, cognitive impairment and quality of life in patients with brain metastases after whole-brain radiotherapy}},
    year = {2023},
    journal = {Clinical and Translational Oncology},
    author = {Yao, Senbang and Zuo, He and Li, Wen and Cai, Yinlian and Zhang, Qianqian and Pang, Lulian and Jing, Yanyan and Yin, Xiangxiang and Cheng, Huaidong},
    number = {1},
    month = {1},
    pages = {207--217},
    volume = {25},
    publisher = {Springer Science and Business Media Deutschland GmbH},
    url = {https://link.springer.com/article/10.1007/s12094-022-02927-3},
    doi = {10.1007/S12094-022-02927-3/FIGURES/6},
    issn = {16993055},
    pmid = {36038750}
}

@article{Kushnirsky2015TheRadiosurgery,
    title = {{The history of stereotactic radiosurgery}},
    year = {2015},
    journal = {Principles and Practice of Stereotactic Radiosurgery},
    author = {Kushnirsky, Marina and Patil, Vaibhav and Schulder, Michael},
    month = {1},
    pages = {3--10},
    publisher = {Springer New York},
    url = {https://link.springer.com/chapter/10.1007/978-1-4614-8363-2_1},
    isbn = {9781461483632},
    doi = {10.1007/978-1-4614-8363-2{\_}1/COVER}
}

@article{VanGrinsven2021TheMeta-Analysis,
    title = {{The Impact of Stereotactic or Whole Brain Radiotherapy on Neurocognitive Functioning in Adult Patients with Brain Metastases: A Systematic Review and Meta-Analysis}},
    year = {2021},
    journal = {Oncology Research and Treatment},
    author = {Van Grinsven, Eva Elisabeth and Nagtegaal, Steven H.J. and Verhoeff, Joost J.C. and Van Zandvoort, Martine J.E.},
    number = {11},
    month = {11},
    pages = {622--636},
    volume = {44},
    publisher = {Karger Publishers},
    url = {https://www.karger.com/Article/FullText/518848 https://www.karger.com/Article/Abstract/518848},
    doi = {10.1159/000518848},
    issn = {2296-5270},
    pmid = {34482312}
}

@article{Kim2023The2023.1,
    title = {{The Korean Society for Neuro-Oncology (KSNO) Guideline for the Management of Brain Tumor Patients During the Crisis Period: A Consensus Survey About Specific Clinical Scenarios (Version 2023.1)}},
    year = {2023},
    journal = {Brain tumor research and treatment},
    author = {Kim, Min-Sung and Go, Se-Il and Wee, Chan Woo and Lee, Min Ho and Kang, Seok-Gu and Go, Kyeong-O and Kwon, Sae Min and Kim, Woohyun and Dho, Yun-Sik and Park, Sung-Hye and Seo, Youngbeom and Song, Sang Woo and Ahn, Stephen and Oh, Hyuk-Jin and Yoon, Hong In and Lee, Sea-Won and Lee, Joo Ho and Cho, Kyung Rae and Choi, Jung Won and Hong, Je Beom and Hwang, Kihwan and Park, Chul-Kee and Lim, Do Hoon},
    number = {2},
    pages = {133},
    volume = {11},
    publisher = {Brain Tumor Res Treat},
    url = {https://pubmed.ncbi.nlm.nih.gov/37151155/},
    doi = {10.14791/BTRT.2023.0010},
    issn = {2288-2405},
    pmid = {37151155}
}

@article{Salvestrini2022TheyAIRO,
    title = {{The role of feature-based radiomics for predicting response and radiation injury after stereotactic radiation therapy for brain metastases: A critical review by the Young Group of the Italian Association of Radiotherapy and Clinical Oncology (yAIRO)}},
    year = {2022},
    journal = {Translational Oncology},
    author = {Salvestrini, Viola and Greco, Carlo and Guerini, Andrea Emanuele and Longo, Silvia and Nardone, Valerio and Boldrini, Luca and Desideri, Isacco and De Felice, Francesca},
    number = {1},
    month = {1},
    pages = {101275},
    volume = {15},
    publisher = {Elsevier},
    doi = {10.1016/J.TRANON.2021.101275},
    issn = {1936-5233}
}

@article{Hasan2014TheMetastases,
    title = {{The role of whole-brain radiation therapy after stereotactic radiation surgery for brain metastases}},
    year = {2014},
    journal = {Practical radiation oncology},
    author = {Hasan, Shaakir and Shah, Ashish H. and Bregy, Amade and Albert, Trevine and Markoe, Arnold and Stoyanova, Radka and Thambuswamy, Michael and Komotar, Ricardo J.},
    number = {5},
    month = {9},
    pages = {306--315},
    volume = {4},
    publisher = {Pract Radiat Oncol},
    url = {https://pubmed.ncbi.nlm.nih.gov/25194099/},
    doi = {10.1016/J.PRRO.2013.09.006},
    issn = {1879-8519},
    pmid = {25194099}
}

@article{Vogelbaum2022TreatmentGuideline,
    title = {{Treatment for Brain Metastases: ASCO-SNO-ASTRO Guideline}},
    year = {2022},
    journal = {Journal of Clinical Oncology},
    author = {Vogelbaum, Michael A. and Brown, Paul D. and Messersmith, Hans and Brastianos, Priscilla K. and Burri, Stuart and Cahill, Dan and Dunn, Ian F. and Gaspar, Laurie E. and Gatson, Na Tosha N. and Gondi, Vinai and Jordan, Justin T. and Lassman, Andrew B. and Maues, Julia and Mohile, Nimish and Redjal, Navid and Stevens, Glen and Sulman, Erik and van den Bent, Martin and Wallace, H. James and Weinberg, Jeffrey S. and Zadeh, Gelareh and Schiff, David},
    number = {5},
    month = {2},
    pages = {492--516},
    volume = {40},
    publisher = {American Society of Clinical Oncology},
    doi = {10.1200/JCO.21.02314},
    issn = {15277755},
    pmid = {34932393}
}

@article{Lin2015TreatmentMetastases,
    title = {{Treatment of Brain Metastases}},
    year = {2015},
    journal = {Journal of Clinical Oncology},
    author = {Lin, Xuling and DeAngelis, Lisa M.},
    number = {30},
    month = {10},
    pages = {3475},
    volume = {33},
    publisher = {American Society of Clinical Oncology},
    url = {/pmc/articles/PMC5087313/ /pmc/articles/PMC5087313/?report=abstract https://www.ncbi.nlm.nih.gov/pmc/articles/PMC5087313/},
    doi = {10.1200/JCO.2015.60.9503},
    issn = {15277755},
    pmid = {26282648}
}

@article{Patil2017WholeMetastases,
    title = {{Whole brain radiation therapy (WBRT) alone versus WBRT and radiosurgery for the treatment of brain metastases}},
    year = {2017},
    journal = {The Cochrane database of systematic reviews},
    author = {Patil, Chirag G. and Pricola, Katie and Sarmiento, J. Manuel and Garg, Sachin K. and Bryant, Andrew and Black, Keith L.},
    number = {9},
    month = {9},
    volume = {9},
    publisher = {Cochrane Database Syst Rev},
    url = {https://pubmed.ncbi.nlm.nih.gov/28945270/},
    doi = {10.1002/14651858.CD006121.PUB4},
    issn = {1469-493X},
    pmid = {28945270}
}

@article{Khan2019WholeFactors,
    title = {{Whole Brain Radiation Therapy Plus Stereotactic Radiosurgery in the Treatment of Brain Metastases Leading to Improved Survival in Patients With Favorable Prognostic Factors}},
    year = {2019},
    journal = {Frontiers in Oncology},
    author = {Khan, Muhammad and Lin, Jie and Liao, Guixiang and Tian, Yunhong and Liang, Yingying and Li, Rong and Liu, Mengzhong and Yuan, Yawei},
    number = {MAR},
    pages = {205},
    volume = {9},
    publisher = {Frontiers Media SA},
    url = {/pmc/articles/PMC6449627/ /pmc/articles/PMC6449627/?report=abstract https://www.ncbi.nlm.nih.gov/pmc/articles/PMC6449627/},
    doi = {10.3389/FONC.2019.00205},
    issn = {2234943X},
    pmid = {30984624}
}

@article{McTyre2013WholeMetastasis,
    title = {{Whole brain radiotherapy for brain metastasis}},
    year = {2013},
    journal = {Surgical Neurology International},
    author = {McTyre, Emory and Scott, Jacob and Chinnaiyan, Prakash},
    number = {Suppl 4},
    month = {10},
    pages = {S236},
    volume = {4},
    publisher = {Scientific Scholar},
    url = {/pmc/articles/PMC3656558/ /pmc/articles/PMC3656558/?report=abstract https://www.ncbi.nlm.nih.gov/pmc/articles/PMC3656558/},
    doi = {10.4103/2152-7806.111301},
    issn = {21527806},
    pmid = {23717795}
}

@article{Brown2018Whole-BrainRevolution,
    title = {{Whole-Brain Radiotherapy for Brain Metastases: Evolution or Revolution?}},
    year = {2018},
    journal = {Journal of clinical oncology : official journal of the American Society of Clinical Oncology},
    author = {Brown, Paul D. and Ahluwalia, Manmeet S. and Khan, Osaama H. and Asher, Anthony L. and Wefel, Jeffrey S. and Gondi, Vinai},
    number = {5},
    month = {2},
    pages = {483--491},
    volume = {36},
    publisher = {J Clin Oncol},
    url = {https://pubmed.ncbi.nlm.nih.gov/29272161/},
    doi = {10.1200/JCO.2017.75.9589},
    issn = {1527-7755},
    pmid = {29272161}
}

@article{Koo2021Whole-BrainStudy,
    title = {{Whole-Brain Radiotherapy vs. Localized Radiotherapy after Resection of Brain Metastases in the Era of Targeted Therapy: A Retrospective Study}},
    year = {2021},
    journal = {Cancers},
    author = {Koo, Jaho and Roh, Tae Hoon and Lee, Sang Ryul and Heo, Jaesung and Oh, Young Taek and Kim, Se Hyuk},
    number = {18},
    month = {9},
    volume = {13},
    publisher = {Cancers (Basel)},
    url = {https://pubmed.ncbi.nlm.nih.gov/34572938/},
    doi = {10.3390/CANCERS13184711},
    issn = {2072-6694},
    pmid = {34572938}
}

\end{document}